\documentclass[pdflatex,sn-mathphys-num]{sn-jnl}

\usepackage{geometry}
\usepackage{subcaption}
\usepackage{graphicx}%
\usepackage{multirow}%
\usepackage{amsmath,amssymb,amsfonts}%
\usepackage{amsthm}%
\usepackage{mathrsfs}%
\usepackage[title]{appendix}%
\usepackage{xcolor}%
\usepackage{textcomp}%
\usepackage{manyfoot}%
\usepackage{booktabs}%
\usepackage{algorithm}%
\usepackage{algorithmicx}%
\usepackage{algpseudocode}%
\usepackage{listings}%

\theoremstyle{thmstyleone}%
\theoremstyle{thmstyletwo}%

\theoremstyle{thmstylethree}%

\usepackage{appendix}
\begin{document}

\title[Article Title]{A model of early word acquisition based on realistic-scale audiovisual naming events}


\author[1]{\fnm{Khazar} \sur{Khorrami}}\email{khazar.khorrami@tuni.fi}

\author*[1]{\fnm{Okko} \sur{Räsänen}}\email{okko.rasanen@tuni.fi}

\affil[1]{\orgdiv{Unit of Computing Sciences}, \orgname{Tampere University}, \orgaddress{\street{P.O. Box 553, FI-33101}, \city{Tampere},\country{ Finland}}}

%

\abstract{Infants gradually learn to parse continuous speech into words and connect names with objects, yet the mechanisms behind development of early word perception skills remain unknown. We studied the extent to which early words can be acquired through statistical learning from regularities in audiovisual sensory input. We simulated word learning in infants up to 12 months of age in a realistic setting, using a model that solely learns from statistical regularities in unannotated raw speech and pixel-level visual input. Crucially, the quantity of object naming events was carefully designed to match that accessible to infants of comparable ages. Results show that the model effectively learns to recognize words and associate them with corresponding visual objects, with a vocabulary growth rate comparable to that observed in infants. The findings support the viability of general statistical learning for early word perception, demonstrating how learning can operate without assuming any prior linguistic capabilities. }



\maketitle

\section{Introduction}\label{sec1}
As they grow, infants gradually acquire understanding of their native language without direct supervision. By the age of 6 months, infants' perception has already attuned to native language phonetic contrasts \cite{polka1994developmental} and they show first signs of word comprehension \cite{tincoff1999some,tincoff2012six,bergelson20126} and familiar word identification \cite{jusczyk1995infants, jusczyk1999beginnings, carbajal2021meta}. By 12 months, they already recognize dozens of words \cite{frank2017wordbank}. During this learning process, the infants must learn to parse the speech stream into words and to associate the words with their referential meanings in the external world. 

\begin{figure*}[!t]%
\centering
\includegraphics[width=1\columnwidth]{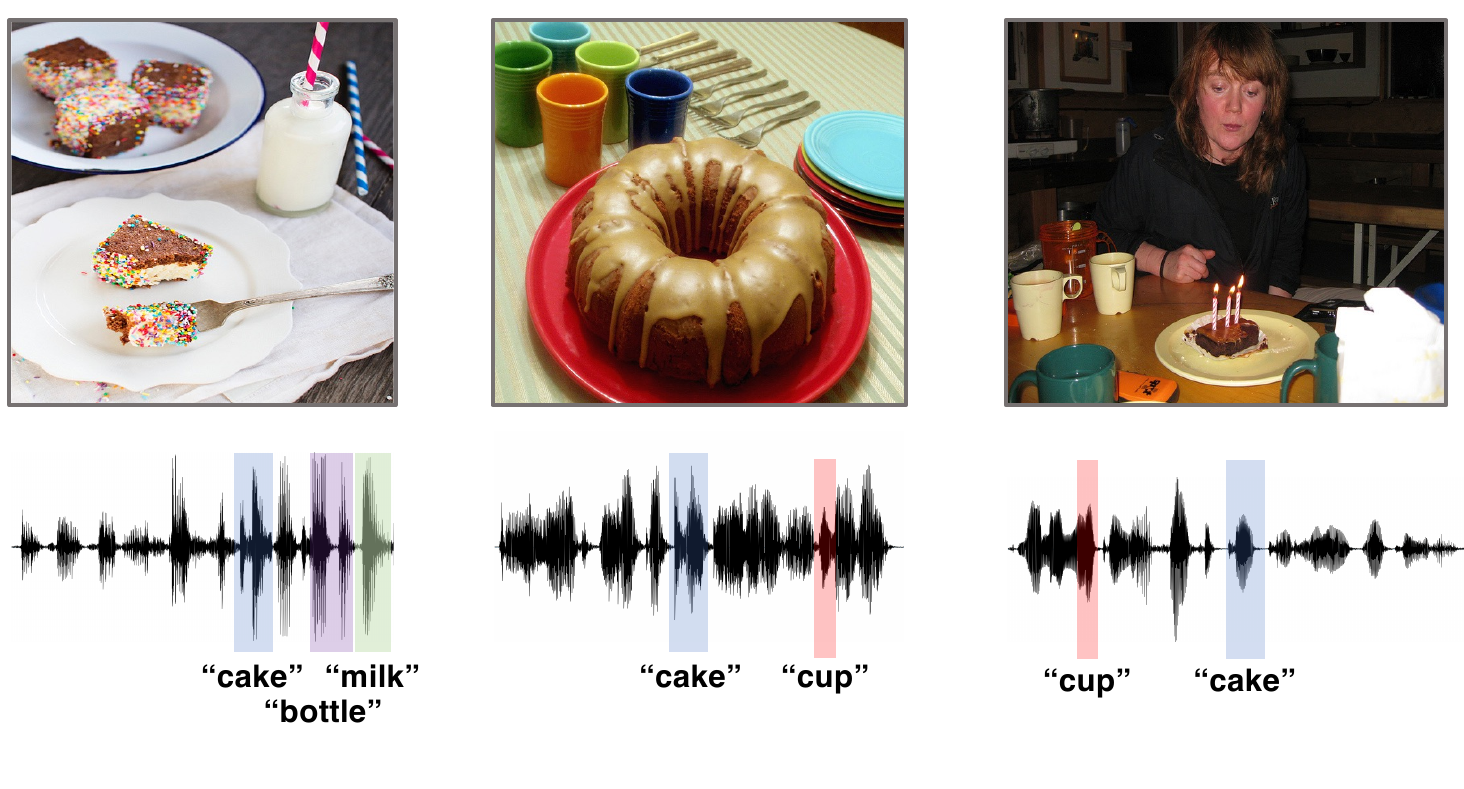}
\caption{\textbf{An illustration of the language learning challenges encountered in realistic settings.} Speech is heard in a particular visual context and the learner has to learn to identify words from running speech, to extract visual objects from the scenes, and then to relate correct words with correct objects by accumulating evidence across multiple individually ambiguous naming events and scenes. This is complicated by the fact that there are no universal cues to word boundaries and each spoken realization of a word has a different acoustic waveform. Utterances also consist of many words that can refer to many different objects in the visual scene, and where only some of the spoken words (highlighted with colors) may be relevant with respect to the current scene. Moreover, only some aspects of the images are named in the utterances, and visual objects of a category also look different across different situations. Finally, one word can refer to many entities of the same visual category (e.g., "cake" in the left panel), and several words can refer to the same physical entity (“bottle” and “milk” for a “bottle of milk” in the left panel). Images and waveforms were extracted from SpokenCOCO dataset \cite{hsu2021text}.}
\label{fig0}
\end{figure*}

From a cognitive perspective, the discovery of words and word-meaning mappings is a task far from trivial: acoustic speech is a continuous and complex signal without transparent linguistic structure (see, e.g., \cite{MOORE2024105694,kuhl2004early}), and there is substantial ambiguity in how individual words embedded in larger utterances are related to specific objects and events in the visual scene, also known as referential ambiguity \cite{quine1960}. Fig. \ref{fig0} illustrates referential ambiguity in everyday life situations between speech and its visual references through some examples. Despite years of extensive research, it remains unclear how infants manage to solve the learning problem, which innate mechanisms or cognitive skills are required, and what can be learned from the data they receive from the external environment. (see, e.g., \cite{gervain2010speech}, for a review).

One prominent approach to understanding early language learning is the so-called \textit{statistical learning hypothesis} (e.g., \citep{saffran1996statistical, saffran2003statistical, maye2002infant, swingley2005statistical}), according to which infants acquire at least some aspects of language by detecting and absorbing statistical regularities\footnote{Sometimes also referred to as distributional learning.}, or patterns, in the sensory input they receive. By doing so, infants can gradually develop capability to identify word boundaries from speech \cite{saffran1996statistical}, but also to relate spoken names to their referents in the external world by tracking co-occurrence statistics of heard words and seen objects when they happen across multiple situations, also known as cross-situational learning \cite{smith2008infants}. However, the extent that statistical learning alone can explain the learning process is challenging to explore with empirical research, as it is difficult to isolate statistical learning from other potential mechanisms and developing cognitive capabilities, such as innate knowledge or reasoning skills.

Complementary to empirical research, computational modeling provides a means to study the language acquisition process in a systematic manner while controlling for the language inputs, learning mechanisms, and outcome measures \citep{dupoux2018cognitive, lavechin2022reverse, lavechin2024modeling, alishahi2017encoding, chrupala2017representations, merkx_thesis, merkx2023modelling, rasanen2015joint, rasanen2015unsupervised,rasanen2019computational,khorrami2021can}. Several existing modeling studies have investigated different aspects of word acquisition. For instance, models of audiovisual cross-situational learning (e.g. \cite{yurovsky2015,yu2012}) have explored how infants might relate words to their visual referents, and how grounding of speech to visual modality may help to segment the continuous speech into words instead of the word segmentation being a prior requirement for the referential grounding \cite{rasanen2015joint}. 

Recent modeling studies with self-supervised models that process unannotated raw speech data have demonstrated successful acquisition of word discrimination skills while using realistic-scale raw speech input in comparison to the amount of input heard by infants during their first couple of years of life \cite{lavechin2022can}. Moreover, visually-grounded speech processing models that learn by aligning between incoming speech signal and concurrent visual scene, have also demonstrated successful self-supervised learning of word forms and word meaning patterns (e.g., \cite{harwath2018jointly,chrupala2022visually}).

However, the amount of audiovisual learning data used in the existing visually-grounded modeling studies has consisted of hundreds of thousands of utterances and speech-related images \citep{khorrami2021can, merkx2023modelling, havard2019word, khorrami2021evaluation, harwath2018jointly}. Research with child-centered at-home video recordings shows that this greatly exceeds the number of audiovisual naming events that real infants experience in their everyday lives \cite{clerkin2019everyday,clerkin2022real}. Since the effectiveness of general statistical learning highly depends on the quantity of data available to the model, it remains an open question whether audiovisual statistical learning from raw acoustic speech and real-world visual data is sufficiently powerful with the relatively fewer audiovisual naming events that infants encounter especially in their first year of age \cite{clerkin2022real}. This is especially pertinent considering the substantial referential ambiguity present in everyday communicative situations, both auditory (several words per utterance) and visually (multiple concurrent visual objects or events in the visual field) \cite{quine1960}.

Recently, \cite{vong2024grounded} investigated word learning grounded in visual data using video recordings from a head-mounted camera worn by a single child from 6 to 25 months of age, thereby capturing the real-life experiences of the child through her eyes.  By using text transcripts as the representation for speech input, they demonstrated that a certain degree of correct word-referent mappings can be achieved through generic associative learning networks even when the visual data is noisy and collected in the wild. However, the use of text transcripts in \cite{vong2024grounded} simplifies the learning problem by substituting continuous and variable speech input with already segmented invariant text-based word forms, thereby bypassing a major part of the learning problem. In reality, word segmentation and categorization is in itself a substantial learning challenge for infants.

In this work, we used computational modeling to investigate whether audiovisual statistical learning from raw speech and visual input scales down to the real-world statistics of infant language experiences from birth to 12 months of age. Our inquiry extends beyond assessing the learner's ability to solve correct word-to-referent mappings, as we also determine the learner's capacity to parse incoming speech into useful linguistic representations that can be associated with their meanings. To study this, we designed a special infant-scale audiovisual dataset using spoken utterances and photographs of everyday scenes from SpokenCOCO dataset \cite{hsu2021text}, where we carefully ensured that the amount of object naming events corresponds to to those reported in experimental studies with infants \cite{clerkin2019everyday,clerkin2022real}. 
Building upon recent advances in unsupervised machine learning, we then employed a combination of unsupervised representation learning (acquisition of higher-order representations from data) and associative learning (learning the relationships between different modalities) as a computational model of an "infant statistical learner".

As a consequence, our simulation closely mirrors the infant language learning process in three key aspects: 1) the learner model receives unnanotated data as input in the form of raw speech audio and pixel-level images, 2) it learns in an unsupervised manner without human guidance, only by discovering statistical regularities within data, and 3) the learning operates on a data scale comparable to what is accessible to real infants. We simulated learning at different checkpoints corresponding to audiovisual input to infants of different ages between 6 and 12 months, and tested the model's knowledge in acoustic word form recognition and in word-to-meaning mappings using a novel benchmark inspired by lab experiments applied for evaluation of infants' word comprehension skills. Our findings demonstrate that word discrimination and word meanings are learnable even with the relatively few naming events available to infant learners, resulting in vocabulary growth compatible with receptive lexicon sizes of real infants of comparable age. Importantly, this learning occurs without strong linguistic priors or innate inductive biases in the learning process.

\begin{figure*}[!t]%
\centering
\includegraphics[width=1\columnwidth]{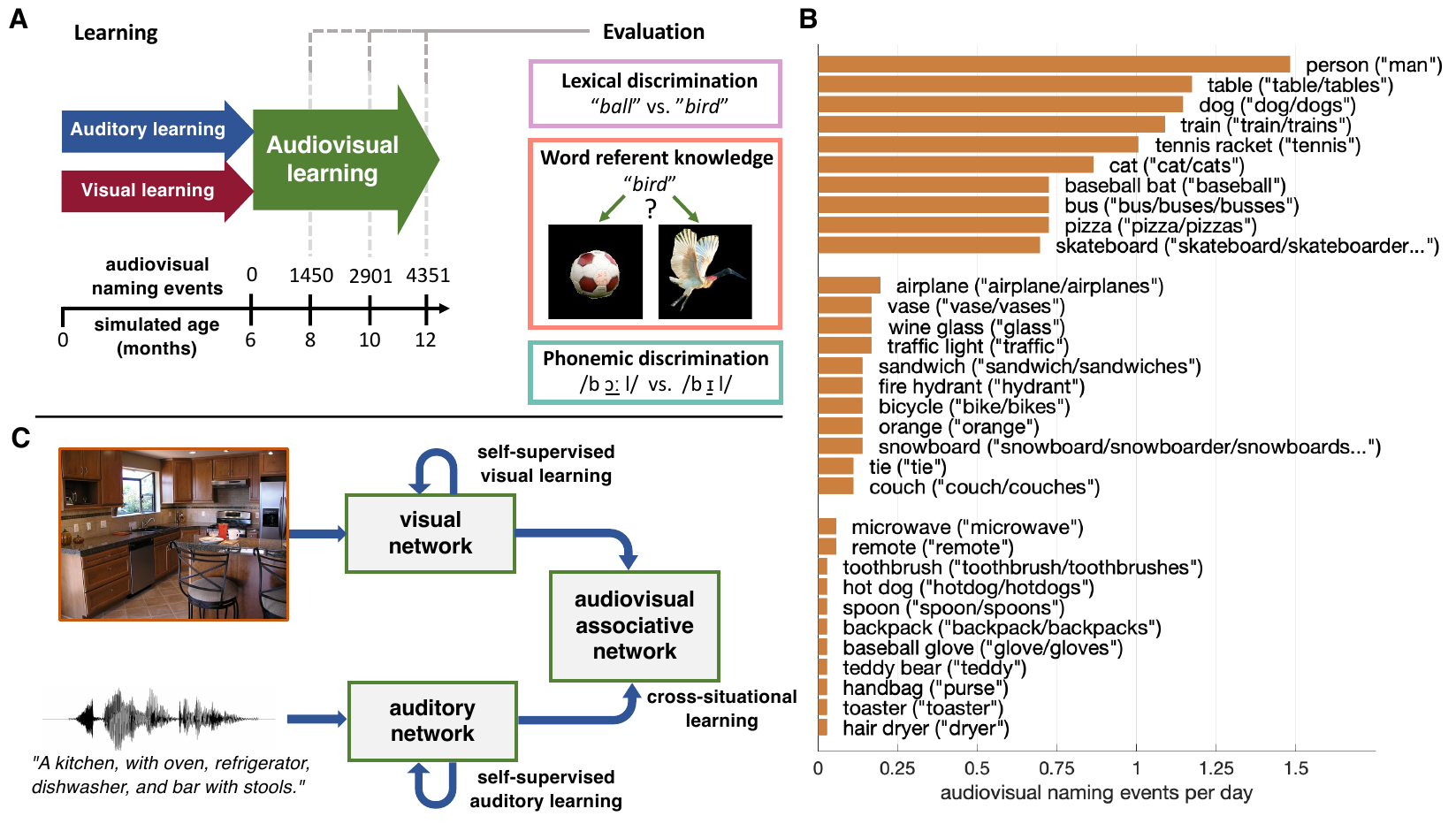}
\caption{\textbf{Modeling approach}. \textbf{A}, Illustration of the modeled timeline in terms of simulated infant months and total number of referentially relevant naming events (across all word types). 
\textbf{B}, Illustration of frequencies at which visual objects and words co-occur in the audiovisual input, with statistics derived from \cite{clerkin2019everyday,clerkin2022real} but adapted to visual object categories of COCO dataset \cite{lin2014microsoft}. Examples of the highest, medium, and the lowest frequency words are shown. The sets of spoken words used to refer to each of the visual categories are shown in parentheses. \textbf{C}, High-level overview of the computational model. Visual and auditory inputs are initially processed in separate encoder networks that first learn in a self-supervised manner without data labels or multimodal influence. After 6 months, cross-modal learning starts to operate in an audiovisual associative network using concurrent sights and sounds together with continued learning in the auditory encoder.}
\label{fig1}
\end{figure*}

\section{Materials and Methods}

\subsection{Overview of the modeling approach}
Fig. \ref{fig1} illustrates the overall modeling approach. We used a statistical learner model to simulate word learning from 0 to 12 months of infant age assuming an initial 6 months of auditory-only and visual-only learning followed by 2-, 4-, or 6-months of the associative learning between speech and concurring visual input, simulating 8-, 10-, 12-month-old infants, respectively (Fig. \ref{fig1}A). This was motivated by the assumption that the auditory and visual systems must develop sufficiently robust representations before being able to associate the modalities with each other (e.g., development of visual constancy \cite{spelke1995}). Furthermore, around six months of age, the infants' developing motor skills start to support more active head control and visual exploration of the surroundings \cite{adolph_2016}, potentially influencing the extent of audio-visual exposures, especially when named objects are not naturally within infants' field of view.

To simulate a realistic amount of auditory and audiovisual learning between the ages of 0 and 12 months, we designed a model training dataset based on empirical statistics of infant language input. For the initial 6 months of auditory training, we created a subset of read speech clips sampled from widely used read speech datasets. This subset was scaled down to an estimated realistic level, based on experimental studies that report the amount of speech per hour to which children are exposed across different families and cultures \cite{bunce2020cross}.

For the training data of our audiovisual associative network, we designed a dataset of image-speech pairs, ensuring that the number of co-occurring word-object pairs closely matched infants' real-life audiovisual language experiences. This design aimed to simulate infants' exposure to object naming events as accurately as possible (Fig. \ref{fig1}B, see also Fig. \ref{fig1}A for the total number of applied audiovisual naming events at each simulated age). For this purpose, we selected a subsample of image-speech pairs from SpokenCOCO \cite{hsu2021text}, a collection of images from everyday life scenes paired with spoken utterances describing the scene contents and events. 

For the data selection process, we employed the object and word occurrence statistics from \cite{clerkin2022real} and \cite{clerkin2019everyday}, which report the rate of naming events for a set of names from infants' early vocabulary. A naming event refers to occasions when an object is in the infant's field of view at the same time that the object is named by adults. The statistics in \cite{clerkin2022real} and \cite{clerkin2019everyday} are derived from head-mounted camera recordings of infants aged 7 to 11 months (mean age of 9 months) during their mealtimes. We applied the same statistics to our simulation data (SpokenCOCO) by substituting the original word and object identities with those present in SpokenCOCO. This resulted in daily naming rates for different visual categories (Fig. \ref{fig1}B), which were then extrapolated for the 2, 4, or 6 months of learning (Fig. \ref{fig1}A). As a reference, we also created a dataset for 4 months of audiovisual learning (simulating 10-month-old babies) where the naming frequency of all categories was set to a uniform distribution, at the same time ensuring that the total number of naming events corresponded to the real-world distribution. This was done to explore whether the non-uniform frequencies of object naming provide a learning advantage (or disadvantage) over the uniform case. 

Note that the distributional characteristics of object names in speech are highly context-dependent (see, e.g., \cite{montag2018quantity} and a related discussion in \cite{clerkin2022real}). In this study, we simply assume that the statistics from one specific scenario (mealtime) represent learning of the specific object names for that scenario. Thereby, the study aimed at testing for the model's capability to learn word meanings from this particular setting, providing a lower bound for potential vocabulary growth that can be then extrapolated across other daily learning scenarios and contexts encountered by infants.

As our model of a statistical learner, we utilized a neural network -based computational model adapted from \citep{peng2022word}. The model is capable of two types of statistical learning: discovery of statistical regularities within individual modalities (auditory or visual processing) and discovery of regularities (associations) between visual and spoken signals that are co-occurring across many situations (Fig. \ref{fig1}C). In the auditory branch, the network learns from raw audio data in a self-supervised manner by attempting to predict temporal segments within the audio stream \citep{baevski2020wav2vec}. In the visual side, a network that processes pixel-level visual inputs is used to simulate visual perception, and it was also trained in a self-supervised manner prior to the present experiments (as explained in \cite{caron2021emerging}). To model cross-situational audiovisual learning, the auditory and visual processing branches are joined together in an audiovisual associative network (similar to \citep{merkx2023modelling,khorrami2021can,harwath2018jointly,chrupala2022visually, peng2022word}). The cross-modal network operates on the outputs of the modality-specific networks with the aim of associating concurrent audiovisual inputs with each other while dissociating non-concurrent inputs. In order to solve the learning task, the model has to learn higher-order representations of the data that support effective inference of yet-to-be-seen input. Thus, the model can be viewed as a practical algorithmic implementation of the statistical learning theory. Moreover, since the model essentially operates by minimizing predictive uncertainty within or across sensory modalities, it also aligns with the idea of mammalian neocortex as a general-purpose sensorimotor prediction machine (e.g., \cite{friston2010})

During the learning process, the model observed speech, images, or pairs of speech and images as input, depending on the learning situation. Visual input consisted of photographs illustrating various everyday-life scenes. Speech input consisted of utterances of speech audio waveforms, each consisting of several words. In the case of audiovisual learning, it was assumed that the utterances are contextually related to the contents of the visual scenes. The model received no information about specific visual objects or audio elements in the input, such as words (see Fig. \ref{fig1}C for an example), or how they relate to each other. Instead, it relied on concurrent pairings of photographs and their spoken descriptions in order to discover any potential cross-modal statistical regularities.  As a consequence, the model had to use statistics of the input to infer how the continuous speech stream is structured in terms of constituent words and how these words relate to objects of the visual world. Development of these capabilities was evaluated together with models' capability to recognize phonemic contrasts (sound meaning distinctions) at each of the age checkpoints (Fig. \ref{fig1}A). 

After exposing the model to auditory-only and visual-only learning to simulate early infancy (0-6 months), audiovisual learning was continued from the age of 6 months onwards using 2, 4, or 6 months worth of speech and image pairs. At each checkpoint, corresponding to simulated infant ages of 8, 10, and 12 months, the language capabilities of the model were evaluated in terms of acoustic phoneme discrimination, word-form discrimination, and word meaning comprehension (distinguishing the visual referents of spoken words) and compared to its performance prior to any learning ("0 months") and after learning from speech only ("6 months"). These metrics were chosen, as it is well known that infants' perception of native phonetic categories and word comprehension start to develop already during the first year of life \cite{polka1994developmental,frank2017wordbank}.

The following subsections provide technical details of the training dataset preparation, model architecture, and evaluation metrics. A reader primarily interested in the simulation outcomes  may directly proceed to the results presented in Section 3.

\subsection{Simulating speech and audiovisual exposure}

\subsubsection{Data sources}

For the initial auditory-only learning stage (0-6 months), we sampled spoken utterances from a combination of LibriSpeech \cite{panayotov2015librispeech} and SpokenCOCO corpora \cite{hsu2021text}. For the audiovisual learning stage (6-12 months), we used pairs of images and their descriptive utterances sampled from SpokenCOCO. 

LibriSpeech \cite{panayotov2015librispeech} originally comprises 960 hours of English read speech, derived from audiobooks, and consisting of a diverse set of speakers and reading styles. The SpokenCOCO dataset \cite{hsu2021text} consists of around 123k images from MSCOCO \cite{lin2014microsoft}, along with their spoken descriptions. In MSCOCO, each image is paired with 5 crowd-sourced text captions that describe the image's visual content, and the dataset covers a total of 80 manually defined common object categories derived from 11 super-categories (animals, furniture, vehicles, etc.), with information on the presence of each object category and their bounding boxes in the images (information which is never made available to our learner model). SpokenCOCO provides crowd-sourced read-aloud image descriptions corresponding to the original MSCOCO captions (2353 speakers; 742 hours of speech in total). Following \citep{peng2022self, peng2022word, khorrami2023computational}, we used the standard test split containing 5k images for model validation at training time and considered the remaining 118k images as the training set from which we sampled our realistic-scale data subsets.
From now on, we refer to the combination of MSCOCO images and SpokenCOCO utterances simply as "COCO". While sampling image-speech pairs for the audiovisual training phase (simulates 2-, 4-, and 6-months of audiovisual learning), we used a subset of COCO image-speech pairs whose audio clips were not part of the initial auditory-only learning data.

\subsubsection{Simulation of realistic-scale infant speech input}

As an estimate of the amount of speech infants hear per day, we followed the large-scale cross-cultural child-centered long-form audio recording data from \cite{bunce2020cross}. However, determining the exact amount of speech input was challenging due to high cross-linguistic \cite{bunce2020cross} and individual variability in language exposure, and since we opted not to continue auditory-only learning in parallel with the audiovisual learning from 6 to 12 months due to technical feasibility.

To obtain 6 months' worth of acoustic speech to learn from (0-6 months of age), and assuming an average of 5.8 hours of speech input per day, as derived from Yélî Dnye total exposure to child- and adult-directed speech per hour (Table SM1 in \cite{bunce2020cross}), and assuming 10 waking hours, auditory learning was modeled with 1049 hours of speech. The resulting 1049 h of speech is a somewhat liberal estimate of speech heard by North-American infants by 6 months of age, but within plausible range for, e.g., Yélî Dnye learners \cite{bunce2020cross} assuming 10 hours of daily awake-time. Since we did not simulate auditory learning beyond the 6-months of age, the input estimate also becomes more accurate as the audiovisual learning proceeds from 6 towards 12 months of age. 

The total of 1049 hours of speech audio was randomly sampled from the mixture of LibriSpeech training set (175,892 clips, 602.6 hours) and the SpokenCOCO training set (370,121 clips, 446.4 hours). We opted for a mixture of LibriSpeech and SpokenCOCO in the auditory learning stage since the total duration of speech in SpokenCOCO alone was insufficient to simulate the 6-months of auditory learning. Additionally, incorporating SpokenCOCO during the pretraining phase helps prevent domain mismatch between the initial auditory and subsequent audiovisual learning stages, as opposed to solely sampling from the LibriSpeech dataset.

\subsubsection{Simulation of realistic-scale object naming events to infants}

For the audiovisual input, the statistics of object and word occurrences from \cite{clerkin2022real} and  \cite{clerkin2019everyday} were utilized. Based on manual annotations of the head-mounted camera video recordings, Clerkin and Smith \cite{clerkin2022real} reported the frequency of naming events, the frequency of objects presented in the scene, and the frequency of object-naming co-occurrences across all the mealtime episodes and within individual mealtimes for a list of 89 early learned word categories for which a spoken name co-occurred with the visual referent at least once in the data (see Supplementary Information in \cite{clerkin2022real}). In general, the naming frequencies of the 89 words followed a skewed distribution with a few very frequent words and a long tail of several infrequent ones. \cite{clerkin2022real} also reported an average of 56.1 minutes of mealtime per day in their data. 

Since the annotated object categories present in COCO are different from the early learned word categories reported in \cite{clerkin2022real}, we aligned the two datasets by sorting the object categories in each dataset from the most frequent to the least frequent and substituting each of the 80 (out of 89) most frequent early learned word categories from \cite{clerkin2022real} with the 80 categories of COCO to obtain realistic-scale object naming frequencies for COCO visual objects.

The number of required object-word pairs per day was then obtained by multiplying the object-specific naming frequencies (now applicable to COCO objects/words) with the average daily mealtime duration, followed by multiplication with an average co-occurrence likelihood of a visual referent to be present during the naming event. This likelihood was set to 50\% based on the statistics available in \cite{clerkin2022real} and \cite{clerkin2019everyday} (i.e., in 50\% of the naming events the corresponding visual referent was also present in the scenario): 

\begin{equation}
\begin{split}
\text{daily object naming rate} = &\ \text{object-specific naming rate (per hour)} \\
&\times \text{average daily mealtime duration (hours)} \\
&\times \text{average object-word co-occurrence likelihood.}
\end{split}
\end{equation}

Finally, the daily audiovisual naming rates were multiplied with the number of days per simulated interval (60, 120 and 180 days) to obtain target counts for audiovisual co-occurrences for each of the 80 categories in the COCO dataset. 


The desired 2-, 4- and 6-month sets of audiovisual data were then created by using the COCO images and corresponding utterances and matching them with the given target statistics. Note that individual images/utterances typically contained more than one visual object/word at a time, and all these were considered towards the total number of co-occurrences. Together with numerous other (non-target) words and photograph contents, the data also contains substantial referential ambiguity (cf., \cite{quine1960}). Fig. \ref{fig1}A shows the resulting total counts for audiovisual naming events (number of concurrent word/object occurrences) and speech hours, whereas Fig. \ref{fig1}B illustrates the resulting audiovisual naming rates for a subset of the categories. 

Since the labels of the visual categories in COCO do not always match with the words used to describe the images (e.g., category \textit{person} described with a word \textit{"man"}), each visual object category was paired with a list of semantically similar and phonologically minimally distinct words. For majority of the objects, this was either the singular or both the singular and plural form of the most frequent word. In some cases, the person using the target object in the image was also used, such as "snowboarder" for "snowboard". All occurrences of the listed words were counted towards the naming counts of the respective objects when preparing the training data (see Fig. \ref{fig1}C for examples; see Appendix C for full word lists). 

Subsequently, the selection of image-utterance pairs for each age subset was executed using a heuristic approach. First, we identified the number of (object, word) samples that were needed for each of the 80 categories in each of the subsets. Next, using the ground-truth image annotations and utterance transcripts, an initial pool of candidate image-utterance pairs was chosen from the COCO train set. From this pool, image-utterance pairs were then incrementally and automatically sampled to obtain a correct number of (object, word) naming events for each of the 80 categories.  Note that one image might contain several target objects and each utterance one or more object names. Moreover, repetitions of object names representing a single category exist in the data (e.g., "a \textit{snowboarder} on a hill with other \textit{snowboarders} in the distance"). We counted all the word repetitions and multiple naming occurrences in the same image-utterance pairs towards the target counts. This process was continued until the maximum number of required samples for each category was reached, simultaneously ensuring that the limit for each category was not exceeded. 

In the simulations, we first trained the model with the auditory learning, followed by training with 2-, 4-, or 6-months worth of audiovisual learning. This allowed us to simulate word learning from from 6 to 12 months of infant age. The resulting word-referent co-occurrence frequencies corresponded to 2--89, 4--178, and 6--267 co-occurrences at 8, 10, and 12 months, respectively.  Additionally, we defined a so-called uniform subset worth of 4 months audio-visual learning by setting the frequency of all concept categories to the mean frequency (=37) over all the 80 categories. The purpose was to study how the distribution of observed naming events affects the learning outcomes compared to the Zipfian distribution of the real naming statistics when the total number of naming events was the same in both conditions (see, e.g., \cite{kurumada2013zipfian}). For the sake of conceptual simplicity, we did not simulate auditory-only learning from non-referential speech beyond the first 6 months.

\subsection{Architecture of the learner model}

We employed the so-called VG-W2V2 neural network architecture as our statistical learner model, initially introduced in \cite{peng2022word} and \cite{peng2022self} and also used as a statistical learner model in \cite{khorrami2023simultaneous}. 
The model is composed of three key components (see Fig. \ref{fig1}C for an overview): auditory and visual encoder blocks that can be trained in a self-supervised (unsupervised) manner independently of each other, and an audio-visual learning block, which maps data from auditory and visual modalities to a shared semantic space, and which is trained using a contrastive cross-modal loss function that operates based on concurrency of spoken utterances and visual images (weakly supervised training). 

\subsubsection{Visual encoder}

On the visual side, the model adapts the image representations obtained from DINO \cite{caron2021emerging}, "self \textbf{di}stillation with \textbf{no} labels", an unsupervised visual representation learning model. DINO follows the architecture of a Vision Transformer (ViT) \cite{dosovitskiy2020image} model and integrates self-supervised learning and knowledge distillation. 
Inputs to the visual processing pipeline consist of RGB images. In ViT \cite{dosovitskiy2020image}, each image is divided into a grid of non-overlapping patches. The array is then unrolled and treated as a sequence. This sequence undergoes processing through a fully connected layer, akin to the word embedding layer in text processing models. The resulting patch embeddings are subsequently input into a transformer block. The overall content of the entire image is ultimately captured by the so-called CLS-token of the final transformer layer, denoted as CLS-I. The DINO model learns by promoting the similarity of different representations derived from the same image while maintaining distinctiveness across different images. The training process involves knowledge transfer from a teacher network to a student network, where the teacher network is constructed from past iterations of the student network. To initiate the training, random perturbations of an image are generated using various data augmentation techniques, such as flipping. Subsequently, the images undergo cropping, with global crops covering more than 50 percent of the image and local crops covering less than 50 percent. The primary objective is to ensure that the student and teacher networks, which share the same architecture (self-distillation), yield identical representations for two different versions of an image, given that both networks receive input derived from the same original image. The training is executed through a cross-entropy loss function applied between the outputs of the student and the teacher networks. Throughout the training process, the student network is directly trained, while the teacher network is constructed from past iterations of the student network, utilizing an exponential moving average (see \cite{caron2021emerging} for a complete DINO description).

Due to computational feasibility considerations, we used a pretrained visual encoder. In accordance with \cite{peng2022word}, we initialized the weights of the image encoder using DINO-ViT small \cite{caron2021emerging}, employing a patch size of 8x8, and pretrained on ImageNet dataset \cite{russakovsky2015imagenet}. The architecture includes an initial linear embedding layer, a positional trainable embedding layer, and a transformer stack with 12 layers, featuring hidden layer dimensionality of 384 with 6 attention heads. The CLS-I token of the final transformer layer was used as the output of the visual encoder.

\subsubsection{Audio encoder}

The first stage of the auditory encoder consists of wav2vec 2.0 \cite{baevski2020wav2vec} architecture. During the self-supervised statistical learning mode, the network's task is to predict its own latent representations for temporal sections of speech masked from the model itself. For this task, the network must learn to use contextual information derived from the non-masked temporal context of the masked sections. The procedure, can be explained as follows: The input utterance, presented as a sequence of short audio frames, is first processed by a 5-layer convolutional neural network (CNN) encoder (512 units in each layer) that maps the signal frames into latent representations occurring every 10-ms. These latents then undergo masking in random temporal segments. The unmasked segments are fed to a stack of 12 transformer layers (768 dim units each) that provide a contextualized representation of the input and a prediction for the masked latents using the CLS-token of the final transformer layer. The network is trained using a contrastive loss, where the optimization objective is to correctly identify the identities of the masked latents amidst distractors randomly sampled from the same utterance. In practice, the latents-to-predict are vector-quantized to categorical entities and the training algorithm seeks to minimize the distance between masked speech representations and quantized versions of their corresponding ground-truth compared to the random distractor tokens. To encourage diverse use of the vector-quantized elements, a diversity loss is incorporated into the loss function. Following the  original study \cite{baevski2020wav2vec}, we combined the masking ($loss_{\text{AUD,R}}$) and diversity ($loss_{\text{AUD,D}}$) losses with a ratio of 1:0.1, denoting their sum as $loss_{\text{AUD}}$.  For a more comprehensive understanding of the training methodology employed in the wav2vec 2.0 model, please refer to the original study by \cite{baevski2020wav2vec}.

\subsubsection{Audio-visual associative network}

To connect the wav2vec 2.0 -based auditory encoder to the visual side, the 8th transformer layer of wav2vec was connected to a following speech encoder block, ResDAVEnet \cite{harwath2019learning}, assigned only for audiovisual training pipeline. ResDAVEnet is a stack of CNN and pooling layers that perform temporal down-sampling of the input. The output of the ResDAVEnet block is passed to a self-attention layer, and the so-called CLS-A token from the this self-attention layer serves as the output speech representation from the auditory encoder.

In the cross-modal associative module of VG-W2V2, the image (CLS-I token) and speech (CLS-A token) representations from the speech and image encoders are merged together. The CLS-I and CLS-A embedding vectors undergo separate projections through 2-layer multi-layer perceptrons (MLPs) that output 2048-dimensional embedding vector representations for both of the modalities. The training employs a masked and marginalized InfoNCE loss  \cite{ilharco2019large} that tries to maximize the similarity between these embeddings from concurrent utterance-image pairs while ensuring dissimilarity of the embeddings from non-concurrent inputs, and where similarity is measured in terms of dot product of the embedding vectors. The InfoNCE of the audiovisual learning pipeline is denoted here as $loss_{\text{AV}}$. Conceptually, the audiovisual network learns to gradually associate semantically relevant elements of spoken utterances to the related image contents (see, e.g., \cite{merkx_thesis, havard2019word,khorrami2021evaluation} for analysis). The result of the dot product of the embeddings was also used in our experiments as the \textit{semantic similarity score} for arbitrary pairs of speech and visual inputs (see below).  

Following \cite{khorrami2023simultaneous}, the audiovisual module was trained by integrating the $loss_{\text{AV}}$ and $loss_{\text{AUD}}$ with a weight $\alpha$ as $loss_{\text{total}} = \alpha  loss_{\text{AV}} + (1-\alpha) loss_{\text{AUD}}$.  Purely auditory learning was conducted using the $loss_{\text{AUD}}$ only.  Note that the CNN encoder and the first 8 transformer layers of the speech encoder are shared between the self-supervised auditory learning and audiovisual learning networks, and also are thereby jointly optimized for the two tasks during audiovisual learning. For a more detailed understanding of the model, please refer to the original VG-w2v2 work presented at \cite{peng2022word} and \cite{peng2022self}.

\subsection{Model training}

For the purely auditory learning simulating the age range from 0 to 6 months, the model was trained with $loss_{AUD}$ for 100 epochs, utilizing a batch size of 22 and a maximum input sequence length of 15 seconds. Validation loss was monitored on a separate held-out section of LibriSpeech, and we saved the model from the epoch with the lowest validation loss. The training took approximately 24 days on 4x V100 GPUs. Since the auditory training with wav2vec 2.0 mechanism faced some initialization issues, likely due to computational constraints that necessitated substantially smaller minibatch size than that used by \cite{baevski2020wav2vec}, we bootstrapped the training by first training a lighter model with only 8 transformer layers in the audio encoder instead of the full 12 for a few epochs. Then the weights from the "pre-trained" CNN block were transferred to the full architecture, the update rate of the CNN layers was set to 0.1, and regular training followed as explained above. 

The audiovisual learner models were trained to simulate infants of age 8, 10, and 12 months using the complete training pipeline of the VG-W2V2 model, employing the $loss_{total}$. Each age bin was trained with the corresponding set of speech-image pairs specifically designed for the 2-, 4-, and 6-months of audiovisual training, as explained earlier. We set $\alpha=0.5$ in $loss_{total}$ for all audiovisual learning experiments to have equal emphasis on auditory self-supervised and audiovisual associative learning during the audiovisual phase. Each age-specific model was initialized using the audio encoder pretrained weights from the 6 months of auditory learning. Each of the audiovisual models was trained for 100 epochs with a batch size of 64. The collective training duration for all audiovisual models was approximately 15 hours, utilizing 4x V100 GPUs. We validated the models every 10 epochs, and the best-performing model was selected and saved based on its audiovisual recall@10 retrieval score (see, e.g., \cite{chrupala2022visually}), measured using the held-out validation set of COCO. In both auditory learning and audiovisual learning, BertAdam optimizer \cite{kenton2019bert} was used with an initial learning rate of 10-4, with a 0.1 warm-up fraction, and then a linear decay towards the end of the training.

\subsection{Evaluation of linguistic capabilities}

Phoneme and word discrimination capabilities were evaluated by analyzing the capability of the auditory encoder hidden layers to discriminate linguistic entities from each other. For this, the representations from the 12 transformer layers of the auditory encoder block were tested separately, and the result from the best-performing layer is always reported as it encapsulates the model's internal knowledge, pertinent to the respective test (see also e.g., \cite{dunbar2022self, peng2022self, khorrami2023computational, khorrami2023simultaneous})). This can also be interpreted as the availability of the related linguistic knowledge for other processing in the "mind" of the learner. The proficiency of the model in comprehending word meanings was evaluated using the semantic similarity score of pairs of audio and images, obtained from the final audio and visual embedding layers.

\subsubsection{Phonemic discrimination} 
Phonemic discrimination was tested using the across-speaker ABX test by \cite{schatz2013evaluating} and using the Zero Resource Speech challenge implementation of the test \cite{dunbar2022self}. The test is a standard approach to evaluate phonemic discrimination in computational models (e.g., \citep{schatz2021early, lavechin2022can, khorrami2023computational}). 
In essence, the test evaluates the input representations' (here: hidden layer activations') ability to differentiate between English minimal pair words, gauging the degree of similarity between two instances of "bit" spoken by different speakers compared to a minimal pair like "bat." The resulting phonemic discrimination error rate describes how often the representations of minimal pair words ("bit" vs. "bat") are closer to each other than two repetitions of the same word ("bit" vs. "bit"), with zero error rate corresponding to perfect phonemic distinctions across all tested comparisons. The evaluation data consists of all possible minimal pairs in the clean-audio development set of the LibriSpeech corpus, comprising a total duration of approximately five hours of speech.

\subsubsection{Word discrimination} Capability of the model to discriminate acoustic word forms from each other was evaluated using CDI-Lextest from \cite{khorrami2023computational}. The test data comprises 89 word types from CDI  North American English short form used to assess infants' receptive vocabularies (\cite{CDI}; data obtained from \cite{frank2017wordbank}). Each of the types has been synthesized into isolated words of speech audio using Google TTS, using 10 different voices, and in speaking styles, resulting in 20 tokens per type and 1780 word tokens in total. 

CDI-Lextest examines how the corresponding model representations encode word identity information (see \cite{chrupala2022visually,lavechin2022reverse} for motivation). During the test, isolated words are fed to the model, and the resulting hidden layer activations of all the tokens are compared within and across word types using cosine distance. The output score (0--100\%; higher is better) of the CDI-Lextest describes how well the model categorizes early infant vocabulary words into distinct perceptual categories by measuring how well the different tokens of the same word type are clustered together in the representation space, where 0\% indicates that given any word token, none of the nearest 19 other tokens are of the same word type, whereas score of 100\% means that, given any token, the 19 most similar tokens are always of the same word type. Consequently, the test evaluates how effectively tokens of each word type cluster together in the representation space while minimizing confusions across different word types. In the test, there is an option to use time-dependent representations from the model being tested, or just one representation vector per isolated word. In the present experiments, the latter option was used (see \cite{khorrami2023computational} for motivation).

\subsubsection{Word meanings} Word meanings were tested in a forced-choice task where the model was exposed to isolated words together with several alternative pictures of visual objects, and the model's judgement of semantic relatedness of the word and the correct image was compared to those of incorrect pairings. For this purpose, the semantic similarity scores from the audiovisual module were employed and a new so-called COCO-Semtest was developed. The test makes use of a dataset of isolated words and visual objects applying a total of 80 different common words and visual categories from the SpokenCOCO. For each of the 80 object categories in COCO dataset, 20 different images consisting of the target object were collected. Using the bounding box annotations of the dataset, we masked all other contents of the image by setting image RGB pixel values to 0 and leaving only the target object visible. In addition, for each of the 80 spoken category names (see Fig. \ref{fig1}B), we synthesized 20  variants of the isolated words using 20 different voices and using Microsoft Azure TTS system. This resulted in a total of 1600 images and 1600 spoken words. 

At the testing time, all possible pairs of object images and spoken words were presented to the model and the resulting semantic similarity scores were recorded. Then, the score of a semantically correct word-image pair was compared against all the semantic similarity scores between the same word but images of different object categories (1580 in total). The proportion (0--100\%) of cases where the semantically correct pair received a higher score than the semantically mismatching pair was then recorded reflecting the proportion of cases where the model identifies the correct visual referent in the presence of a competing distractor of another visual category. The process was repeated for all the 1600 word tokens, and the mean of the token-specific proportion scores of a word type is reported as the word meaning score of the category, and the final word meaning score is then reported as the mean across all categories (chance-level = 50\%). The score can be interpreted as an outcome of two-alternative forced-choice task: given an isolated word and a correct and wrong visual referent, what is the probability that the model considers the correct pairing as semantically more relevant (see Fig. 1B for a conceptual illustration).

\section{Results}

\begin{figure}[tp]%
\centering
\begin{subfigure}{0.9\columnwidth}
\centering
\includegraphics[width=\linewidth]{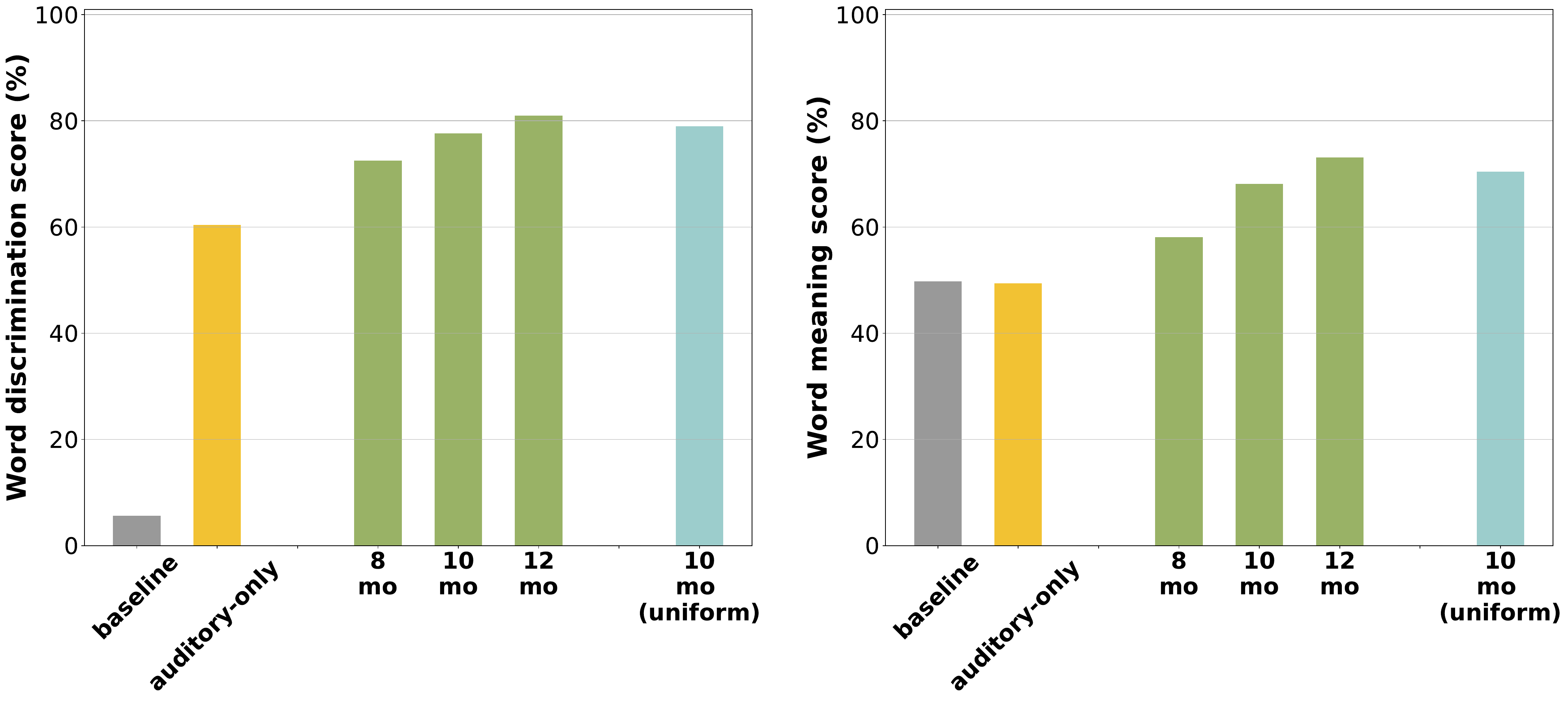}
\caption{Word learning scores of the simulated learner at different ages (in months). Left: word discrimination score. Right: word meaning score. Baseline checkpoint corresponds to a model with randomly initialized parameters. Auditory-only refers to a model that is trained only using auditory learning from speech. The 8, 10, and 12 months bars correspond to the visually-grounded trained models that are each followed by initial auditory-only training. 10 mo (uniform) refers to 10-month checkpoint with uniform object-name co-occurrence frequencies during learning.}
\label{fig:results0}
\end{subfigure}

\vspace{0.4cm} 

\begin{subfigure}{0.99\columnwidth}
\centering
\includegraphics[width=\linewidth]{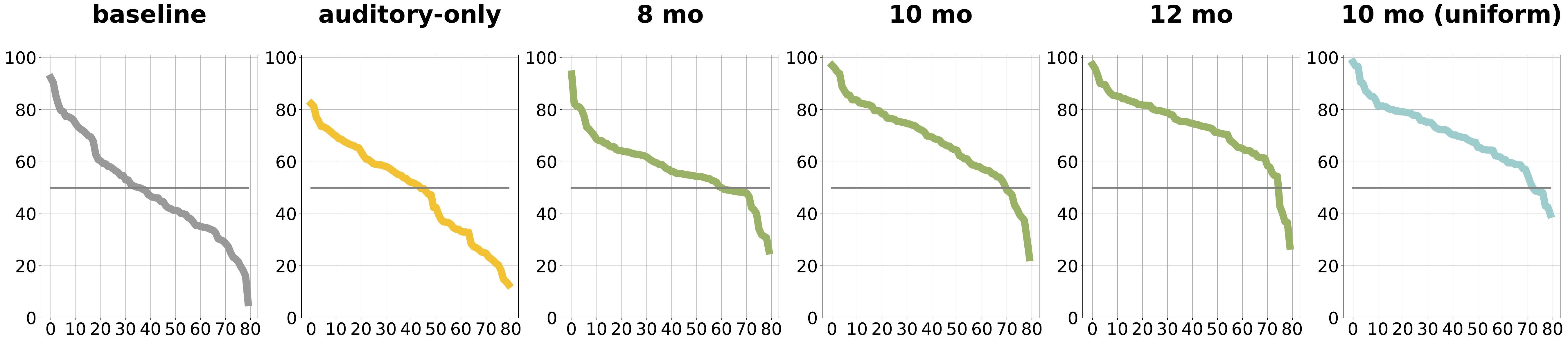}
\caption{The distribution of the word meaning scores over categories for the baseline and the trained models. Please note that the scores are internally sorted for each model and thus the order of the categories on the x-axis may differ across the models. }
\label{fig:results1}
\end{subfigure}

\vspace{0.4cm} 

\begin{subfigure}{0.99\columnwidth}
\centering
\includegraphics[width=\linewidth]{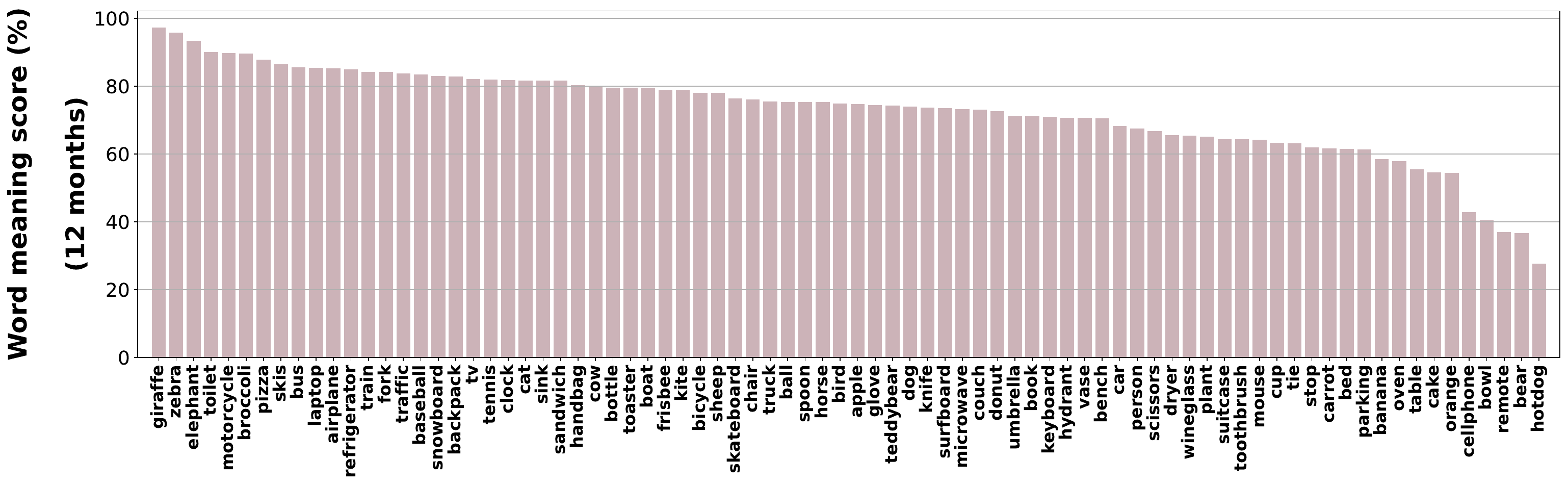}
\caption{The distribution of the word meaning scores over categories for the model of a 12-month-old infant.} 
\label{fig:results2}
\end{subfigure}

\caption{\textbf{Words learning competence in a simulated infant of age 0-12 months-old.} \textbf{a,} Word discrimination and word meaning learning for different trained models. \textbf{b,c,} The distribution of word meaning scores over various object categories. }
\label{figResults}
\end{figure}

Fig. \ref{fig:results0} shows the results for the word discrimination and word meaning scores as a function of simulated learner age, alongside with a baseline model that has not yet observed any audio or audiovisual data (i.e., a model with random initial parameters) and the model that has observed only audio data. 

For word discrimination (Fig. \ref{fig:results0}, left), auditory-only learning already reached a discrimination performance of 60.1\%, which is substantially above chance-level (1.12\%). Moreover, the word discrimination performance further increased in a systematic manner with increasing amounts of audiovisual learning up to a score of 80.7\% at 12 months of age. In the case of uniform word-referent occurrence frequencies across the categories ('10 mo (uniform)' in Fig. \ref{fig:results0}), the model demonstrates a slightly improved word discrimination competence (78.7\%) compared to the scenario with a naturalistic skewed distribution (72.2\%). Overall, the results demonstrate how auditory statistical learning can result in good acoustic word form discrimination skills (see also \cite{lavechin23_interspeech} \cite{lavechin2022can}  for convergent findings) that can be then significantly improved by further audiovisual associative training. It is also noteworthy that the amount of additional speech input during the audiovisual learning (in terms of speech hours) is very minor compared to the initial auditory-only learning in our simulations. This result demonstrates, for the first time, the effectiveness of audiovisual learning as a powerful strategy for acquiring stable word form representations from limited data, even in the presence of substantial referential ambiguity.

The right panel in Fig. \ref{fig:results0} shows the model's capability to associate correct visual referents with spoken words. As expected, the performance is at chance level of 50\% after purely auditory learning. In contrast, the word meaning score systematically increases with increasing audiovisual learning experience, obtaining above-chance level (57.8\%) already at 8 months
and reaching 72.8\% at 12 months of age. Similar to the word discrimination, the word meaning score is slightly higher with the uniform distribution (70.1\%) compared to the natural distribution (67.8\%) at 10 months. 

Fig. \ref{fig:results1} illustrates the distribution of the word meaning scores over categories for the baseline and the trained models, where the scores are taken from individually sorted distributions. Although the order of categories might change across different models, the overall distribution across categories can be compared. The results shows a more uniform distribution over categories for the baseline and auditory-only trained model whereas audio-visually trained models represent a less uniform distribution that is skewed at both ends indicating notable variations in the respective scores. For instance, at  8 months, 19 (out of 80) categories perform at chance while the best-performing category ("giraffe") achieves a high score of 93.7\%. After 10 months of training, the majority of categories show a word meaning score above the chance level (see also Fig. \ref{fig3}), and at 12 months, only 5 out of 80 object categories had a word meaning score that was indifferent from the chance level. Fig. \ref{fig:results2} provides information on the object category names for the distribution of word meaning scores over categories at age of 12 months. Please see the \textit{Appendix B} for a complete category-by-category breakdown of the scores for all the age groups.

In terms of phonemic perception, the initial speech-only learning resulted in good phonemic discrimination competence with an approximately 7.1\%  ABX error rate (with the chance-level being 50\%). The continued audiovisual learning beyond the auditory learning did not result in further improvements in the ABX score, nor did it compromise the performance either. This aligns with the earlier findings on auditory statistical learning as a means to learn high competence in phonemic discrimination in the native language of the learner \cite{lavechin2022can,dunbar2022self}. In fact, when we tested audiovisual learning directly without the preceding auditory learning, the model was unable to learn anything from the small number of audiovisual naming events with the test metrics remaining at chance-level. This highlights the importance of learning statistical patterns of the auditory speech stream, here reflected by the good phonemic discrimination skills, in order to support cross-modal associative learning from a small number of naming events. 

Overall, the results indicate that the learner successfully associated various acoustic word-forms with their visual referents in real images of everyday scenes. This was achieved despite the words and their referents not occurring in isolation during training, but as part of larger utterances or visual scenes.

\subsection{Post-hoc analyses}
Given that some object-word categories were learned better than others, we also performed post-hoc statistical analyses to understand in more detail what could explain the differences in the learning outcomes. More specifically, we tested whether the category-level word meaning scores at the 8-, 10-, and 12-month checkpoints were correlated with the frequencies of the corresponding naming events. In addition, we tested whether the word meaning scores were correlated with the visual object sizes in the training data, parametrized as the proportion of image area that an object covered on average, as visually more dominant objects might be easier for the model to learn.  Table \ref{tabCorr} shows the outcomes of the analyses. The frequency of audiovisual object naming explained some of the variance in the learning outcomes, ranging from Spearman $\rho$ = 0.243 at 8 months to $\rho$ = 0.334 at 12 months. In contrast, visual object size had no effect on the word learning outcomes.

\begin{table}[htbp]
\caption{Spearman correlation coefficients $\rho$ (and p-values) for correlations between object area or object naming frequency and word meaning scores of the objects.}
\centering
\begin{tabular}{|p{0.22\columnwidth}|p{0.1\columnwidth}|p{0.1\columnwidth}|p{0.1\columnwidth}|p{0.11\columnwidth}|} 
\hline
& \textbf{8 mo} & \textbf{10 mo} & \textbf{12 mo} & \textbf{10 mo(u)}   \\
\hline 
 \textbf{Object area} & n.s. & n.s. & n.s. &  n.s  \\
\hline
\textbf{Naming frequency} & 0.243 (p\(=\)0.030) & 0.260 (p\(=\)0.020) & 0.334 (p\(=\)0.002) &  - \\
\hline 
\end{tabular}
\label{tabCorr}
\end{table}

\subsection{Qualitative comparison to infant vocabulary growth}

We also tested how the development of the word comprehension in our model compares to vocabulary growth in infants. Development of infant word comprehension is typically measured using parental questionnaires from the so-called Child Development Inventories (CDI; \cite{CDI}), which consist of checklists for words that an infant understands at a given age. Size of the receptive vocabulary of an infant can then be defined as the total number of words that the infant understands. We used CDI vocabulary data from hundreds of North American English infants, as available in Wordbank database \cite{frank2017wordbank}, and compared the vocabulary sizes of the infants to our model's vocabulary size at different age checkpoints. 

\begin{figure}[tp]%
\centering
\includegraphics[width=0.9\columnwidth]{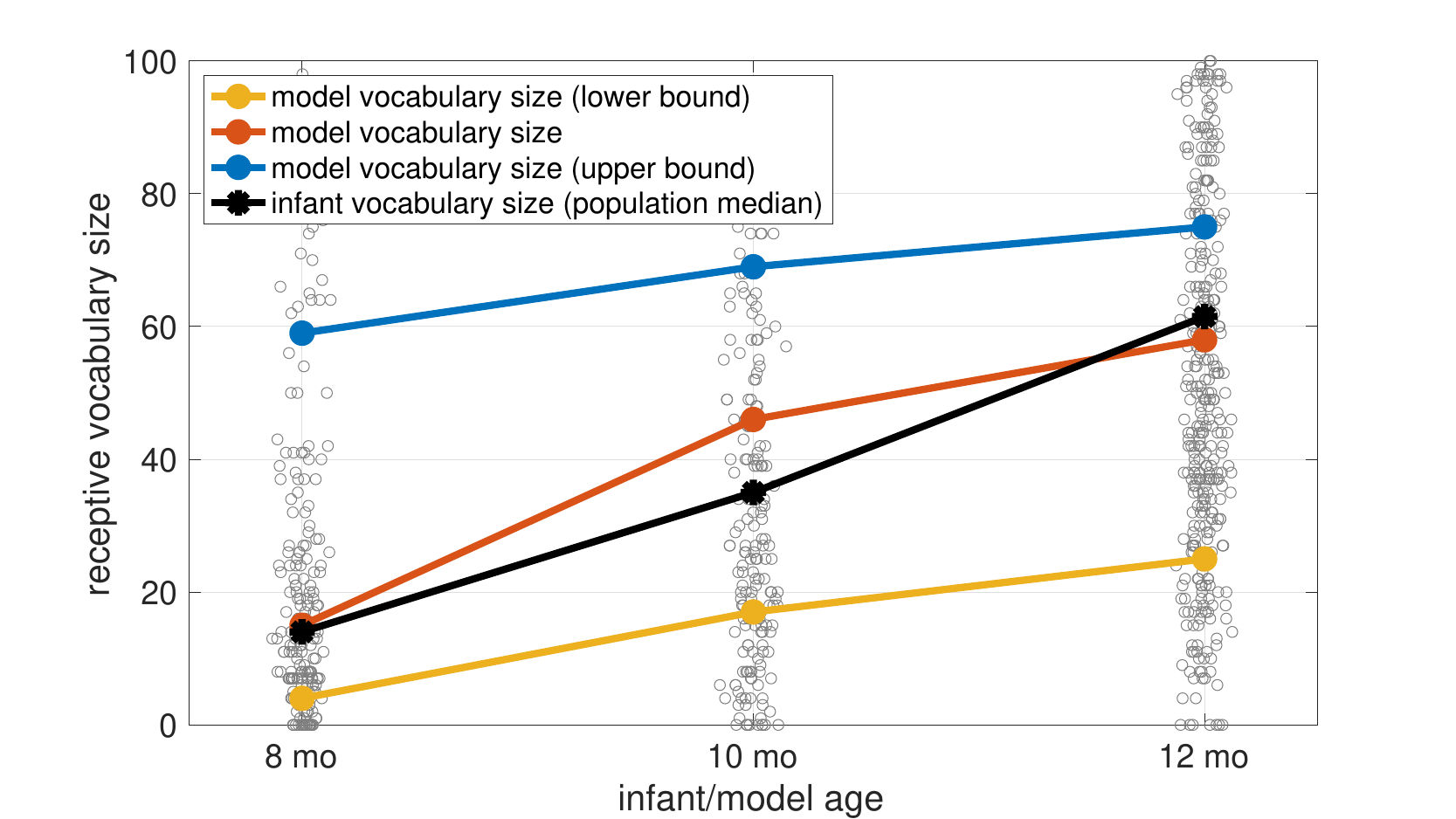}
\caption{\textbf{Vocabulary size of the model and real infants in terms of word comprehension}. Black line denotes median vocabulary size in the infant population and grey circles show vocabulary sizes of individual infants (data from CDI vocabulary norms of North American English-learning infants \cite{frank2017wordbank}). Three different vocabulary size curves are shown for the model, depending on how accurately the model needs to identify the correct visual referent of a spoken word in the two-alternative forced-choice task used for word meaning scores. Blue line: above-chance identification of the correct referent (upper bound for vocabulary size). Red line: correct in 2/3 of the trials. Yellow line: correct in 4/5 of the trials (lower bound).}
\label{fig3}
\end{figure}

Fig. \ref{fig3} shows the infant vocabulary data together with three different vocabulary curves for the model. Each model curve differs in terms of how consistently the model needs to identify the correct visual referent for a word in the two-alternative forced choice task, as there is no generic criterion for defining word as learned. As can be seen from the figure, the number of words that the model understands is in similar range with the vocabulary sizes of infants of the same age. Note that exact one-to-one comparison between the infants and the model is not possible. This is because the infant vocabulary size estimates are based on the parental reports for word understanding using the CDI checklists consisting of hundreds of words, whereas the model vocabulary size was defined based on the 80 words and visual categories tested in the two-alternative forced-choice task. Moreover, parental reports cover vocabulary from different areas of infant life but the present simulation setup only covers learning from one recurring everyday context: the mealtimes. Hence, the figure qualitatively illustrates how human-compatible vocabulary sizes are already achievable by modeling audiovisual statistical learning from a fraction of the waking hours of infants, but where detailed comparison is not possible.

\section{Discussion and Conclusions}
This study set out to test whether generic statistical learning within and across sensory modalities is powerful enough to learn words and word meanings from the relatively few good audiovisual naming events that infants encounter in their daily lives. We defined "good" naming events as those where a word and its visual referent co-occur in the same communicative context (as defined in \cite{clerkin2019everyday,clerkin2022real}), but where substantial referential ambiguity was still present in terms of speech (with many words in the same utterance) and the visual scene (real photographs of everyday scenes with many objects, and as represented by pixel maps). We used a self-supervised computational model as our statistical learner, and matched the properties of the input data available to the learner with empirical estimates of infant language experiences. By doing this, we could assess whether the model succeeds in acquiring language comprehension skills from the real and highly variable sensory input. As a result, we found that the model is able to bootstrap phonetic and lexical learning, including referential word meanings, from the simulated language experience up to 12 months of age. 

As far as we know, this is the first evidence that statistical learning from real audio and visual data at a realistic scale enables acquisition of referential word meanings, while also achieving vocabulary growth that is compatible with what is observed in real
infants. It is important to note that the computational model never received any explicit prior knowledge, expert labels, or feedback regarding linguistic structure of speech or structure of visual scenes.  Moreover, it never observed visual objects or words related to them in isolation, but always embedded into complex scenes and utterances related to the scenes. The model still learned to solve the word segmentation and meaning mapping problems by using self-supervised mechanisms that can be viewed as minimization of predictive uncertainty within and across sensory modalities. By minimizing predictive uncertainty in time or across modalities, the network was forced to learn latent representations that best support the prediction tasks. In case of speech input, this optimization appears to result in emergent internal representations that are highly correlated with theoretical linguistic concepts such as phonemes and words. 

Previous research on auditory and audiovisual statistical learning has already demonstrated a plethora of phenomena that can be modeled with present-type of neural models of visually-grounded speech statistical learners: by learning to align contents of spoken utterances with objects of visual events, self-supervised audiovisual models also implicitly acquire linguistically-motivated representations. The findings include consistent emergence of phonological \cite{chrupala2017representations,alishahi2017encoding}, lexical \cite{chrupala2022visually,merkx2023modelling},  and syllabic \cite{peng2023syllable,khorrami2021can} structure in the hidden layers, and that the emergence of these different levels of representation takes place in parallel during the learning time  \cite{khorrami2023simultaneous} and does not depend on model architecture details \cite{khorrami2021can}. Notably, the emergence of this type of latent linguistic knowledge, dubbed "Latent Language Hypothesis" in \citep{khorrami2021can}, takes place without any prior linguistic information in the model design and without ever aiming to learn such units from the input. As argued in \cite{merkx_thesis}, such findings can be interpreted as evidence for usage-based theory of language acquisition \cite{tomasello2009} in conjunction with statistical learning mechanisms for bootstrapping language acquisition: by hearing speech in various everyday communicative situations, statistical learners may start from holistic meaning-centered utterance representations and gradually learn to decompose them into linguistic constituents, such as individual words or sub-word elements. 

However, earlier literature never tested learning with a realistic amount of audiovisual naming events. Only a recent  modeling study by \cite{vong2024grounded} showed how word-to-meaning mappings can be learned from the audiovisual input available to an infant, aligning with our present findings. Notably, the model in \cite{vong2024grounded} did not include the learning of basic speech perception capabilities, such as word segmentation or phonemic perception, but instead used transcribed speech as input to their model, thereby bypassing the auditory learning problem. When results of that study are combined with the present evidence that the learning also succeeds from real audio, the literature as a whole suggests that a large proportion of the early speech comprehension could be explained in terms of generic statistical learning mechanisms. Yet, as with any computational modeling research, these findings should be viewed as a learnability proof of what is achievable in principle, and where infants may acquire comparable results with somewhat different statistical learning mechanisms.

Evidently, the present study has several limitations that need to be considered regarding the potential generality of the results. These limitations can be addressed in future research as suitable computational setups, including data, models, and corresponding evaluation metrics, become available. First, the auditory input was not real infant-directed speech, but consisted of spoken utterances of read-aloud captions describing images. This choice was motivated due to the lack of a suitable real-world datasets with high-enough audio quality, as the current computational models struggle with acoustic input mixed with environmental noise \cite{lavechin_struggle}, a common feature in real child-centered audio. Moreover, the tests used to probe the model's linguistic competence operate on clean speech, and an acoustic mismatch between the model's learning input and testing input can have a notable impact on the evaluation outcomes. In an ideal case, the audiovisual input would consist of head-mounted camera video from infants (e.g., \cite{bergelson20126,clerkin2022real}) instead of photographs and their spoken descriptions. Yet, a comparable audio quality and scale issue is present with those data, limiting the use of actual audio in models using such data (see, e.g., \cite{vong2024grounded}).

Second, there are differences between our simulation dataset (COCO) and the real-world statistics from \cite{clerkin2022real} that may affect modeling accuracy. A notable distinction is that the data in \cite{clerkin2022real} focuses solely on mealtime, presenting a single context. As a result, a real infant learner is unable to use context to differentiate between object categories. In contrast, the COCO dataset contains numerous scene categories, naturally clustering most objects into separate contexts. For example, a 'car' and 'hydrant' are typically found on the street, while a 'spoon' is frequent inside a home. This contextual diversity might aid the model in recognizing objects and words by incorporating contextual information (see also \cite{roy2015}).  We also extrapolated the real-world statistics obtained from 7-11 months old infants \cite{clerkin2022real} to the entire simulated range of 6--12 months. However, \cite{clerkin2022real} have considered all the objects present in the scene as a visual occurrence, even when the object was not in the current field of view of the infant, while according to the 'object permanence' theory \cite{piaget2013construction}, before 8 months, when an object is out of sight, it is akin to non-existence, potentially impacting assumptions about infants' awareness of visual objects.

Finally, our simulation omits cognitive development and infant-caregiver interaction dynamics, which evolve alongside statistical learning. Factors like memory, attention, and social skills likely influence data reception and processing. Incorporation of such factors into computational models would be important, but are beyond the scope of the present aim: to test the feasibility of statistical learning in itself.

The exact extent that the listed limitations affect the main findings is difficult to estimate. However, we believe that the present simulation setup is not making the learning problem substantially easier compared to real life. Instead, the lack of infant-tailored speech input, lack of infant's point of view visual input, and the lack of social interaction likely make the learning problem more challenging. 

Overall, the results provide strong support for statistical learning as a feasible mechanism to support early language acquisition by showing how word learning can get started without requiring any prior knowledge of how speech is organized (e.g., in terms of phonemes, syllables, or words) and without requiring dedicated learning mechanisms that target specific linguistic structures, such as mechanisms for word segmentation or phoneme categorization.

\section*{Data availability}
All program code, including model architecture, experiment scripts, data split specifications, and the COCO-Semtest, will be made available on GitHub upon acceptance for publication. An archived persistent version of all the program code and associated data will also be made available on Zenodo upon acceptance. CDI-Lextest and ABX test are publicly available from their respective GitHub repositories. LibriSpeech, SpokenCOCO and MSCOCO datasets are publicly available with their respective licenses.

\section*{Acknowledgements}This research was funded by Academy of Finland grants no. 314602 and 345365, and by Kone Foundation grant to L-SCALE -project. The authors wish to acknowledge CSC – IT Center for Science, Finland, for generous computational resources.

\section*{Author contributions}
K.K. and O.R. conceptualized and designed the research. K.K. performed data curation, implemented the model, and conducted all computational experiments. K.K. and O.R. implemented model evaluation metrics and conducted result analyses. K.K. was primarily responsible for writing the paper, with O.R. also contributing to the writing, reviewing, and commenting on the manuscript. K.K. and O.R. collaboratively created the visualizations. O.R. supervised the work and acquired the funding.

\section*{Competing interests}
The authors declare no competing interest.

\section*{Declaration of Generative AI and AI-assisted technologies in the writing process} 

During the preparation of this work the authors used ChatGPT (OpenAI, GPT-3 and GPT-4) in order to improve the manuscript's readability and for grammatical proofreading of the text. After using this tool, the authors reviewed and edited the content as needed and take full responsibility for the content of the publication.

\bibliography{main}


\begin{thebibliography}{69}
\ifx \bisbn   \undefined \def \bisbn  #1{ISBN #1}\fi
\ifx \binits  \undefined \def \binits#1{#1}\fi
\ifx \bauthor  \undefined \def \bauthor#1{#1}\fi
\ifx \batitle  \undefined \def \batitle#1{#1}\fi
\ifx \bjtitle  \undefined \def \bjtitle#1{#1}\fi
\ifx \bvolume  \undefined \def \bvolume#1{\textbf{#1}}\fi
\ifx \byear  \undefined \def \byear#1{#1}\fi
\ifx \bissue  \undefined \def \bissue#1{#1}\fi
\ifx \bfpage  \undefined \def \bfpage#1{#1}\fi
\ifx \blpage  \undefined \def \blpage #1{#1}\fi
\ifx \burl  \undefined \def \burl#1{\textsf{#1}}\fi
\ifx \doiurl  \undefined \def \doiurl#1{\url{https://doi.org/#1}}\fi
\ifx \betal  \undefined \def \betal{\textit{et al.}}\fi
\ifx \binstitute  \undefined \def \binstitute#1{#1}\fi
\ifx \binstitutionaled  \undefined \def \binstitutionaled#1{#1}\fi
\ifx \bctitle  \undefined \def \bctitle#1{#1}\fi
\ifx \beditor  \undefined \def \beditor#1{#1}\fi
\ifx \bpublisher  \undefined \def \bpublisher#1{#1}\fi
\ifx \bbtitle  \undefined \def \bbtitle#1{#1}\fi
\ifx \bedition  \undefined \def \bedition#1{#1}\fi
\ifx \bseriesno  \undefined \def \bseriesno#1{#1}\fi
\ifx \blocation  \undefined \def \blocation#1{#1}\fi
\ifx \bsertitle  \undefined \def \bsertitle#1{#1}\fi
\ifx \bsnm \undefined \def \bsnm#1{#1}\fi
\ifx \bsuffix \undefined \def \bsuffix#1{#1}\fi
\ifx \bparticle \undefined \def \bparticle#1{#1}\fi
\ifx \barticle \undefined \def \barticle#1{#1}\fi
\bibcommenthead
\ifx \bconfdate \undefined \def \bconfdate #1{#1}\fi
\ifx \botherref \undefined \def \botherref #1{#1}\fi
\ifx \url \undefined \def \url#1{\textsf{#1}}\fi
\ifx \bchapter \undefined \def \bchapter#1{#1}\fi
\ifx \bbook \undefined \def \bbook#1{#1}\fi
\ifx \bcomment \undefined \def \bcomment#1{#1}\fi
\ifx \oauthor \undefined \def \oauthor#1{#1}\fi
\ifx \citeauthoryear \undefined \def \citeauthoryear#1{#1}\fi
\ifx \endbibitem  \undefined \def \endbibitem {}\fi
\ifx \bconflocation  \undefined \def \bconflocation#1{#1}\fi
\ifx \arxivurl  \undefined \def \arxivurl#1{\textsf{#1}}\fi
\csname PreBibitemsHook\endcsname

\bibitem[\protect\citeauthoryear{Polka and Werker}{1994}]{polka1994developmental}
\begin{barticle}
\bauthor{\bsnm{Polka}, \binits{L.}},
\bauthor{\bsnm{Werker}, \binits{J.F.}}:
\batitle{Developmental changes in perception of nonnative vowel contrasts}.
\bjtitle{Journal of Experimental Psychology: Human Perception and Performance}
\bvolume{20}(\bissue{2}),
\bfpage{421}
(\byear{1994})
\end{barticle}
\endbibitem

\bibitem[\protect\citeauthoryear{Tincoff and Jusczyk}{1999}]{tincoff1999some}
\begin{barticle}
\bauthor{\bsnm{Tincoff}, \binits{R.}},
\bauthor{\bsnm{Jusczyk}, \binits{P.W.}}:
\batitle{Some beginnings of word comprehension in 6-month-olds}.
\bjtitle{Psychological science}
\bvolume{10}(\bissue{2}),
\bfpage{172}--\blpage{175}
(\byear{1999})
\end{barticle}
\endbibitem

\bibitem[\protect\citeauthoryear{Tincoff and Jusczyk}{2012}]{tincoff2012six}
\begin{barticle}
\bauthor{\bsnm{Tincoff}, \binits{R.}},
\bauthor{\bsnm{Jusczyk}, \binits{P.W.}}:
\batitle{Six-month-olds comprehend words that refer to parts of the body}.
\bjtitle{Infancy}
\bvolume{17}(\bissue{4}),
\bfpage{432}--\blpage{444}
(\byear{2012})
\end{barticle}
\endbibitem

\bibitem[\protect\citeauthoryear{Bergelson and Swingley}{2012}]{bergelson20126}
\begin{barticle}
\bauthor{\bsnm{Bergelson}, \binits{E.}},
\bauthor{\bsnm{Swingley}, \binits{D.}}:
\batitle{At 6--9 months, human infants know the meanings of many common nouns}.
\bjtitle{Proceedings of the National Academy of Sciences}
\bvolume{109}(\bissue{9}),
\bfpage{3253}--\blpage{3258}
(\byear{2012})
\end{barticle}
\endbibitem

\bibitem[\protect\citeauthoryear{Jusczyk and Aslin}{1995}]{jusczyk1995infants}
\begin{barticle}
\bauthor{\bsnm{Jusczyk}, \binits{P.W.}},
\bauthor{\bsnm{Aslin}, \binits{R.N.}}:
\batitle{Infants' detection of the sound patterns of words in fluent speech}.
\bjtitle{Cognitive psychology}
\bvolume{29}(\bissue{1}),
\bfpage{1}--\blpage{23}
(\byear{1995})
\end{barticle}
\endbibitem

\bibitem[\protect\citeauthoryear{Jusczyk et~al.}{1999}]{jusczyk1999beginnings}
\begin{barticle}
\bauthor{\bsnm{Jusczyk}, \binits{P.W.}},
\bauthor{\bsnm{Houston}, \binits{D.M.}},
\bauthor{\bsnm{Newsome}, \binits{M.}}:
\batitle{The beginnings of word segmentation in english-learning infants}.
\bjtitle{Cognitive psychology}
\bvolume{39}(\bissue{3-4}),
\bfpage{159}--\blpage{207}
(\byear{1999})
\end{barticle}
\endbibitem

\bibitem[\protect\citeauthoryear{Carbajal et~al.}{2021}]{carbajal2021meta}
\begin{barticle}
\bauthor{\bsnm{Carbajal}, \binits{M.J.}},
\bauthor{\bsnm{Peperkamp}, \binits{S.}},
\bauthor{\bsnm{Tsuji}, \binits{S.}}:
\batitle{A meta-analysis of infants’ word-form recognition}.
\bjtitle{Infancy}
\bvolume{26}(\bissue{3}),
\bfpage{369}--\blpage{387}
(\byear{2021})
\end{barticle}
\endbibitem

\bibitem[\protect\citeauthoryear{Frank et~al.}{2017}]{frank2017wordbank}
\begin{barticle}
\bauthor{\bsnm{Frank}, \binits{M.C.}},
\bauthor{\bsnm{Braginsky}, \binits{M.}},
\bauthor{\bsnm{Yurovsky}, \binits{D.}},
\bauthor{\bsnm{Marchman}, \binits{V.A.}}:
\batitle{Wordbank: An open repository for developmental vocabulary data}.
\bjtitle{Journal of Child Language}
\bvolume{44}(\bissue{3}),
\bfpage{677}--\blpage{694}
(\byear{2017})
\end{barticle}
\endbibitem

\bibitem[\protect\citeauthoryear{Moore and Bergelson}{2024}]{MOORE2024105694}
\begin{barticle}
\bauthor{\bsnm{Moore}, \binits{C.}},
\bauthor{\bsnm{Bergelson}, \binits{E.}}:
\batitle{Wordform variability in infants’ language environment and its effects on early word learning}.
\bjtitle{Cognition}
\bvolume{245},
\bfpage{105694}
(\byear{2024})
\end{barticle}
\endbibitem

\bibitem[\protect\citeauthoryear{Kuhl}{2004}]{kuhl2004early}
\begin{barticle}
\bauthor{\bsnm{Kuhl}, \binits{P.K.}}:
\batitle{Early language acquisition: cracking the speech code}.
\bjtitle{Nature reviews neuroscience}
\bvolume{5}(\bissue{11}),
\bfpage{831}--\blpage{843}
(\byear{2004})
\end{barticle}
\endbibitem

\bibitem[\protect\citeauthoryear{Quine}{1960}]{quine1960}
\begin{botherref}
\oauthor{\bsnm{Quine}, \binits{W.V.O.}}:
Word and object.
MIT Press,
(1960)
\end{botherref}
\endbibitem

\bibitem[\protect\citeauthoryear{Gervain and Mehler}{2010}]{gervain2010speech}
\begin{barticle}
\bauthor{\bsnm{Gervain}, \binits{J.}},
\bauthor{\bsnm{Mehler}, \binits{J.}}:
\batitle{Speech perception and language acquisition in the first year of life}.
\bjtitle{Annual Review of Psychology}
\bvolume{61},
\bfpage{191}--\blpage{218}
(\byear{2010})
\end{barticle}
\endbibitem

\bibitem[\protect\citeauthoryear{Saffran et~al.}{1996}]{saffran1996statistical}
\begin{barticle}
\bauthor{\bsnm{Saffran}, \binits{J.R.}},
\bauthor{\bsnm{Aslin}, \binits{R.N.}},
\bauthor{\bsnm{Newport}, \binits{E.L.}}:
\batitle{Statistical learning by 8-month-old infants}.
\bjtitle{Science}
\bvolume{274}(\bissue{5294}),
\bfpage{1926}--\blpage{1928}
(\byear{1996})
\end{barticle}
\endbibitem

\bibitem[\protect\citeauthoryear{Saffran}{2003}]{saffran2003statistical}
\begin{barticle}
\bauthor{\bsnm{Saffran}, \binits{J.R.}}:
\batitle{Statistical language learning: Mechanisms and constraints}.
\bjtitle{Current Directions in Psychological Science}
\bvolume{12}(\bissue{4}),
\bfpage{110}--\blpage{114}
(\byear{2003})
\end{barticle}
\endbibitem

\bibitem[\protect\citeauthoryear{Maye et~al.}{2002}]{maye2002infant}
\begin{barticle}
\bauthor{\bsnm{Maye}, \binits{J.}},
\bauthor{\bsnm{Werker}, \binits{J.F.}},
\bauthor{\bsnm{Gerken}, \binits{L.}}:
\batitle{Infant sensitivity to distributional information can affect phonetic discrimination}.
\bjtitle{Cognition}
\bvolume{82}(\bissue{3}),
\bfpage{101}--\blpage{111}
(\byear{2002})
\end{barticle}
\endbibitem

\bibitem[\protect\citeauthoryear{Swingley}{2005}]{swingley2005statistical}
\begin{barticle}
\bauthor{\bsnm{Swingley}, \binits{D.}}:
\batitle{Statistical clustering and the contents of the infant vocabulary}.
\bjtitle{Cognitive psychology}
\bvolume{50}(\bissue{1}),
\bfpage{86}--\blpage{132}
(\byear{2005})
\end{barticle}
\endbibitem

\bibitem[\protect\citeauthoryear{Smith and Yu}{2008}]{smith2008infants}
\begin{barticle}
\bauthor{\bsnm{Smith}, \binits{L.}},
\bauthor{\bsnm{Yu}, \binits{C.}}:
\batitle{Infants rapidly learn word-referent mappings via cross-situational statistics}.
\bjtitle{Cognition}
\bvolume{106}(\bissue{3}),
\bfpage{1558}--\blpage{1568}
(\byear{2008})
\end{barticle}
\endbibitem

\bibitem[\protect\citeauthoryear{Hsu et~al.}{2021}]{hsu2021text}
\begin{bchapter}
\bauthor{\bsnm{Hsu}, \binits{W.-N.}},
\bauthor{\bsnm{Harwath}, \binits{D.}},
\bauthor{\bsnm{Miller}, \binits{T.}},
\bauthor{\bsnm{Song}, \binits{C.}},
\bauthor{\bsnm{Glass}, \binits{J.}}:
\bctitle{Text-free image-to-speech synthesis using learned segmental units}.
In: \bbtitle{Proc. 59th Annual Meeting of the Association for Computational Linguistics and the 11th International Joint Conference on Natural Language Processing (Volume 1: Long Papers)},
pp. \bfpage{5284}--\blpage{5300}
(\byear{2021})
\end{bchapter}
\endbibitem

\bibitem[\protect\citeauthoryear{Dupoux}{2018}]{dupoux2018cognitive}
\begin{barticle}
\bauthor{\bsnm{Dupoux}, \binits{E.}}:
\batitle{Cognitive science in the era of artificial intelligence: A roadmap for reverse-engineering the infant language-learner}.
\bjtitle{Cognition}
\bvolume{173},
\bfpage{43}--\blpage{59}
(\byear{2018})
\end{barticle}
\endbibitem

\bibitem[\protect\citeauthoryear{Lavechin et~al.}{2022}]{lavechin2022reverse}
\begin{barticle}
\bauthor{\bsnm{Lavechin}, \binits{M.}},
\bauthor{\bsnm{Seyssel}, \binits{M.}},
\bauthor{\bsnm{Gautheron}, \binits{L.}},
\bauthor{\bsnm{Dupoux}, \binits{E.}},
\bauthor{\bsnm{Cristia}, \binits{A.}}:
\batitle{Reverse engineering language acquisition with child-centered long-form recordings}.
\bjtitle{Annual Review of Linguistics}
\bvolume{8},
\bfpage{389}--\blpage{407}
(\byear{2022})
\end{barticle}
\endbibitem

\bibitem[\protect\citeauthoryear{Lavechin et~al.}{2024}]{lavechin2024modeling}
\begin{barticle}
\bauthor{\bsnm{Lavechin}, \binits{M.}},
\bauthor{\bsnm{Seyssel}, \binits{M.}},
\bauthor{\bsnm{M{\'e}tais}, \binits{M.}},
\bauthor{\bsnm{Metze}, \binits{F.}},
\bauthor{\bsnm{Mohamed}, \binits{A.}},
\bauthor{\bsnm{Bredin}, \binits{H.}},
\bauthor{\bsnm{Dupoux}, \binits{E.}},
\bauthor{\bsnm{Cristia}, \binits{A.}}:
\batitle{Modeling early phonetic acquisition from child-centered audio data}.
\bjtitle{Cognition}
\bvolume{245},
\bfpage{105734}
(\byear{2024})
\end{barticle}
\endbibitem

\bibitem[\protect\citeauthoryear{Alishahi et~al.}{2017}]{alishahi2017encoding}
\begin{bchapter}
\bauthor{\bsnm{Alishahi}, \binits{A.}},
\bauthor{\bsnm{Barking}, \binits{M.}},
\bauthor{\bsnm{Chrupa{\l}a}, \binits{G.}}:
\bctitle{Encoding of phonology in a recurrent neural model of grounded speech}.
In: \bbtitle{{Proc. 21st Conference on Computational Natural Language Learning ({C}o{NLL} 2017)}},
pp. \bfpage{368}--\blpage{378}
(\byear{2017})
\end{bchapter}
\endbibitem

\bibitem[\protect\citeauthoryear{Chrupa{\l}a et~al.}{2017}]{chrupala2017representations}
\begin{bchapter}
\bauthor{\bsnm{Chrupa{\l}a}, \binits{G.}},
\bauthor{\bsnm{Gelderloos}, \binits{L.}},
\bauthor{\bsnm{Alishahi}, \binits{A.}}:
\bctitle{Representations of language in a model of visually grounded speech signal}.
In: \bbtitle{{Proc. 55th Annual Meeting of the Association for Computational Linguistics (Volume 1: Long Papers)}},
pp. \bfpage{613}--\blpage{622}
(\byear{2017})
\end{bchapter}
\endbibitem

\bibitem[\protect\citeauthoryear{Merkx}{2002}]{merkx_thesis}
\begin{botherref}
\oauthor{\bsnm{Merkx}, \binits{D.}}:
Modelling multi-modal language learning: from sentences to words.
Doctoral thesis, Nijmegen: MPI
(2002)
\end{botherref}
\endbibitem

\bibitem[\protect\citeauthoryear{Merkx et~al.}{2023}]{merkx2023modelling}
\begin{barticle}
\bauthor{\bsnm{Merkx}, \binits{D.}},
\bauthor{\bsnm{Scholten}, \binits{S.}},
\bauthor{\bsnm{Frank}, \binits{S.L.}},
\bauthor{\bsnm{Ernestus}, \binits{M.}},
\bauthor{\bsnm{Scharenborg}, \binits{O.}}:
\batitle{Modelling human word learning and recognition using visually grounded speech}.
\bjtitle{Cognitive Computation}
\bvolume{15}(\bissue{1}),
\bfpage{272}--\blpage{288}
(\byear{2023})
\end{barticle}
\endbibitem

\bibitem[\protect\citeauthoryear{R{\"a}s{\"a}nen and Rasilo}{2015}]{rasanen2015joint}
\begin{barticle}
\bauthor{\bsnm{R{\"a}s{\"a}nen}, \binits{O.}},
\bauthor{\bsnm{Rasilo}, \binits{H.}}:
\batitle{A joint model of word segmentation and meaning acquisition through cross-situational learning.}
\bjtitle{Psychological Review}
\bvolume{122}(\bissue{4}),
\bfpage{792}--\blpage{829}
(\byear{2015})
\end{barticle}
\endbibitem

\bibitem[\protect\citeauthoryear{R{\"a}s{\"a}nen et~al.}{2015}]{rasanen2015unsupervised}
\begin{bchapter}
\bauthor{\bsnm{R{\"a}s{\"a}nen}, \binits{O.}},
\bauthor{\bsnm{Doyle}, \binits{G.}},
\bauthor{\bsnm{Frank}, \binits{M.C.}}:
\bctitle{Unsupervised word discovery from speech using automatic segmentation into syllable-like units}.
In: \bbtitle{Sixteenth Annual Conference of the International Speech Communication Association}
(\byear{2015})
\end{bchapter}
\endbibitem

\bibitem[\protect\citeauthoryear{R{\"{a}}s{\"{a}}nen and Khorrami}{2019}]{rasanen2019computational}
\begin{bchapter}
\bauthor{\bsnm{R{\"{a}}s{\"{a}}nen}, \binits{O.}},
\bauthor{\bsnm{Khorrami}, \binits{K.}}:
\bctitle{A computational model of early language acquisition from audiovisual experiences of young infants}.
In: \bbtitle{{20th Annual Conference of the International Speech Communication Association (Interspeech 2019)}},
pp. \bfpage{3594}--\blpage{3598}
(\byear{2019})
\end{bchapter}
\endbibitem

\bibitem[\protect\citeauthoryear{Khorrami and R{\"a}s{\"a}nen}{2021}]{khorrami2021can}
\begin{barticle}
\bauthor{\bsnm{Khorrami}, \binits{K.}},
\bauthor{\bsnm{R{\"a}s{\"a}nen}, \binits{O.}}:
\batitle{Can phones, syllables, and words emerge as side-products of cross-situational audiovisual learning? {-- A} computational investigation}.
\bjtitle{Language Development Research}
\bvolume{1},
\bfpage{123}--\blpage{191}
(\byear{2021})
\end{barticle}
\endbibitem

\bibitem[\protect\citeauthoryear{Yurovsky and Frank}{2015}]{yurovsky2015}
\begin{barticle}
\bauthor{\bsnm{Yurovsky}, \binits{D.}},
\bauthor{\bsnm{Frank}, \binits{M.C.}}:
\batitle{An integrative account of constraints on cross-situational learning}.
\bjtitle{Cognition}
\bvolume{145},
\bfpage{53}--\blpage{62}
(\byear{2015})
\end{barticle}
\endbibitem

\bibitem[\protect\citeauthoryear{Yu and Smith}{2012}]{yu2012}
\begin{barticle}
\bauthor{\bsnm{Yu}, \binits{C.}},
\bauthor{\bsnm{Smith}, \binits{L.}}:
\batitle{Modeling cross-situational word–referent learning: Prior questions}.
\bjtitle{Psychological Review}
\bvolume{119}(\bissue{1}),
\bfpage{21}--\blpage{39}
(\byear{2012})
\end{barticle}
\endbibitem

\bibitem[\protect\citeauthoryear{de~Seyssel et~al.}{2023}]{lavechin2022can}
\begin{barticle}
\bauthor{\bsnm{Seyssel}, \binits{M.}},
\bauthor{\bsnm{Lavechin}, \binits{M.}},
\bauthor{\bsnm{Dupoux}, \binits{E.}}:
\batitle{Realistic and broad-scope learning simulations: first results and challenges}.
\bjtitle{Journal of Child Language}
\bvolume{50},
\bfpage{1}--\blpage{24}
(\byear{2023})
\end{barticle}
\endbibitem

\bibitem[\protect\citeauthoryear{Harwath et~al.}{2018}]{harwath2018jointly}
\begin{bchapter}
\bauthor{\bsnm{Harwath}, \binits{D.}},
\bauthor{\bsnm{Recasens}, \binits{A.}},
\bauthor{\bsnm{Sur{\'\i}s}, \binits{D.}},
\bauthor{\bsnm{Chuang}, \binits{G.}},
\bauthor{\bsnm{Torralba}, \binits{A.}},
\bauthor{\bsnm{Glass}, \binits{J.}}:
\bctitle{Jointly discovering visual objects and spoken words from raw sensory input}.
In: \bbtitle{Proc. European Conference on Computer Vision (ECCV-2018)},
pp. \bfpage{649}--\blpage{665}
(\byear{2018})
\end{bchapter}
\endbibitem

\bibitem[\protect\citeauthoryear{Chrupa{\l}a}{2022}]{chrupala2022visually}
\begin{barticle}
\bauthor{\bsnm{Chrupa{\l}a}, \binits{G.}}:
\batitle{Visually grounded models of spoken language: A survey of datasets, architectures and evaluation techniques}.
\bjtitle{Journal of Artificial Intelligence Research}
\bvolume{73},
\bfpage{673}--\blpage{707}
(\byear{2022})
\end{barticle}
\endbibitem

\bibitem[\protect\citeauthoryear{Havard et~al.}{2019}]{havard2019word}
\begin{bchapter}
\bauthor{\bsnm{Havard}, \binits{W.N.}},
\bauthor{\bsnm{Chevrot}, \binits{J.-P.}},
\bauthor{\bsnm{Besacier}, \binits{L.}}:
\bctitle{Word recognition, competition, and activation in a model of visually grounded speech}.
In: \bbtitle{Proceedings of the 23rd Conference on Computational Natural Language Learning (CoNLL)},
pp. \bfpage{339}--\blpage{348}
(\byear{2019})
\end{bchapter}
\endbibitem

\bibitem[\protect\citeauthoryear{Khorrami and R{\"a}s{\"a}nen}{2021}]{khorrami2021evaluation}
\begin{bchapter}
\bauthor{\bsnm{Khorrami}, \binits{K.}},
\bauthor{\bsnm{R{\"a}s{\"a}nen}, \binits{O.}}:
\bctitle{Evaluation of audio-visual alignments in visually grounded speech models}.
In: \bbtitle{Proc. Annual Conference of the International Speech Communication Association (Interspeech-2021)},
pp. \bfpage{1231}--\blpage{1235}
(\byear{2021}).
\bcomment{International Speech Communication Association}
\end{bchapter}
\endbibitem

\bibitem[\protect\citeauthoryear{Clerkin and Smith}{2019}]{clerkin2019everyday}
\begin{bchapter}
\bauthor{\bsnm{Clerkin}, \binits{E.M.}},
\bauthor{\bsnm{Smith}, \binits{L.B.}}:
\bctitle{The everyday statistics of objects and their names: How word learning gets its start}.
In: \bbtitle{Proc. Annual Conference of the Cognitive Science Society (CogSci-2019)},
pp. \bfpage{240}--\blpage{246}
(\byear{2019})
\end{bchapter}
\endbibitem

\bibitem[\protect\citeauthoryear{Clerkin and Smith}{2022}]{clerkin2022real}
\begin{barticle}
\bauthor{\bsnm{Clerkin}, \binits{E.M.}},
\bauthor{\bsnm{Smith}, \binits{L.B.}}:
\batitle{Real-world statistics at two timescales and a mechanism for infant learning of object names}.
\bjtitle{Proceedings of the National Academy of Sciences}
\bvolume{119}(\bissue{18}),
\bfpage{2123239119}
(\byear{2022})
\end{barticle}
\endbibitem

\bibitem[\protect\citeauthoryear{Vong et~al.}{2024}]{vong2024grounded}
\begin{barticle}
\bauthor{\bsnm{Vong}, \binits{W.K.}},
\bauthor{\bsnm{Wang}, \binits{W.}},
\bauthor{\bsnm{Orhan}, \binits{A.E.}},
\bauthor{\bsnm{Lake}, \binits{B.M.}}:
\batitle{Grounded language acquisition through the eyes and ears of a single child}.
\bjtitle{Science}
\bvolume{383}(\bissue{6682}),
\bfpage{504}--\blpage{511}
(\byear{2024})
\end{barticle}
\endbibitem

\bibitem[\protect\citeauthoryear{Lin et~al.}{2014}]{lin2014microsoft}
\begin{bchapter}
\bauthor{\bsnm{Lin}, \binits{T.-Y.}},
\bauthor{\bsnm{Maire}, \binits{M.}},
\bauthor{\bsnm{Belongie}, \binits{S.}},
\bauthor{\bsnm{Hays}, \binits{J.}},
\bauthor{\bsnm{Perona}, \binits{P.}},
\bauthor{\bsnm{Ramanan}, \binits{D.}},
\bauthor{\bsnm{Doll{\'a}r}, \binits{P.}},
\bauthor{\bsnm{Zitnick}, \binits{C.L.}}:
\bctitle{Microsoft coco: Common objects in context}.
In: \bbtitle{Proc. European Conference on Computer Vision (ECCV-2014)},
pp. \bfpage{740}--\blpage{755}
(\byear{2014})
\end{bchapter}
\endbibitem

\bibitem[\protect\citeauthoryear{Spelke et~al.}{1995}]{spelke1995}
\begin{bchapter}
\bauthor{\bsnm{Spelke}, \binits{E.S.}},
\bauthor{\bsnm{Vishton}, \binits{P.}},
\bauthor{\bsnm{Hofsten}, \binits{C.}}:
\bctitle{Object perception, object-directed action, and physical knowledge in infancy}.
In: \beditor{\bsnm{Gazzaniga}, \binits{M.S.}} (ed.)
\bbtitle{The Cognitive Neurosciences},
pp. \bfpage{165}--\blpage{179}.
\bpublisher{MIT Press}, \blocation{???}
(\byear{1995})
\end{bchapter}
\endbibitem

\bibitem[\protect\citeauthoryear{Adolph and Franchak}{2016}]{adolph_2016}
\begin{botherref}
\oauthor{\bsnm{Adolph}, \binits{K.E.}},
\oauthor{\bsnm{Franchak}, \binits{J.M.}}:
The development of motor behavior.
WIREs Cognitive Science
\textbf{e1430}
(2016)
\end{botherref}
\endbibitem

\bibitem[\protect\citeauthoryear{Bunce et~al.}{}]{bunce2020cross}
\begin{botherref}
\oauthor{\bsnm{Bunce}, \binits{J.}},
\oauthor{\bsnm{Soderstrom}, \binits{M.}},
\oauthor{\bsnm{Bergelson}, \binits{E.}},
\oauthor{\bsnm{Rosemberg}, \binits{C.}},
\oauthor{\bsnm{Stein}, \binits{A.}},
\oauthor{\bsnm{Migdalek}, \binits{M.}},
\oauthor{\bsnm{Casillas}, \binits{M.}}, et al.:
A cross-cultural examination of young children’s everyday language experiences.
PsyArXiv pre-print
\end{botherref}
\endbibitem

\bibitem[\protect\citeauthoryear{Montag et~al.}{2018}]{montag2018quantity}
\begin{barticle}
\bauthor{\bsnm{Montag}, \binits{J.L.}},
\bauthor{\bsnm{Jones}, \binits{M.N.}},
\bauthor{\bsnm{Smith}, \binits{L.B.}}:
\batitle{Quantity and diversity: Simulating early word learning environments}.
\bjtitle{Cognitive Science}
\bvolume{42},
\bfpage{375}--\blpage{412}
(\byear{2018})
\end{barticle}
\endbibitem

\bibitem[\protect\citeauthoryear{Peng and {Harwath}}{2022}]{peng2022word}
\begin{bchapter}
\bauthor{\bsnm{Peng}, \binits{P.}},
\bauthor{\bsnm{{Harwath}}, \binits{D.}}:
\bctitle{Word discovery in visually grounded, self-supervised speech models}.
In: \bbtitle{Proc. Annual Conference of the International Speech Communication Association (Interspeech-2022)},
pp. \bfpage{2823}--\blpage{2827}
(\byear{2022})
\end{bchapter}
\endbibitem

\bibitem[\protect\citeauthoryear{Baevski et~al.}{2020}]{baevski2020wav2vec}
\begin{barticle}
\bauthor{\bsnm{Baevski}, \binits{A.}},
\bauthor{\bsnm{Zhou}, \binits{Y.}},
\bauthor{\bsnm{Mohamed}, \binits{A.}},
\bauthor{\bsnm{Auli}, \binits{M.}}:
\batitle{wav2vec 2.0: A framework for self-supervised learning of speech representations}.
\bjtitle{Advances in Neural Information Processing Systems}
\bvolume{33},
\bfpage{12449}--\blpage{12460}
(\byear{2020})
\end{barticle}
\endbibitem

\bibitem[\protect\citeauthoryear{Caron et~al.}{2021}]{caron2021emerging}
\begin{bchapter}
\bauthor{\bsnm{Caron}, \binits{M.}},
\bauthor{\bsnm{Touvron}, \binits{H.}},
\bauthor{\bsnm{Misra}, \binits{I.}},
\bauthor{\bsnm{J{\'e}gou}, \binits{H.}},
\bauthor{\bsnm{Mairal}, \binits{J.}},
\bauthor{\bsnm{Bojanowski}, \binits{P.}},
\bauthor{\bsnm{Joulin}, \binits{A.}}:
\bctitle{Emerging properties in self-supervised vision transformers}.
In: \bbtitle{Proc. IEEE/CVF International Conference on Computer Vision},
pp. \bfpage{9650}--\blpage{9660}
(\byear{2021})
\end{bchapter}
\endbibitem

\bibitem[\protect\citeauthoryear{Friston}{2010}]{friston2010}
\begin{barticle}
\bauthor{\bsnm{Friston}, \binits{K.}}:
\batitle{The free-energy principle: a unified brain theory?}
\bjtitle{Nature Reviews Neuroscience}
\bvolume{11},
\bfpage{127}--\blpage{138}
(\byear{2010})
\end{barticle}
\endbibitem

\bibitem[\protect\citeauthoryear{Panayotov et~al.}{2015}]{panayotov2015librispeech}
\begin{bchapter}
\bauthor{\bsnm{Panayotov}, \binits{V.}},
\bauthor{\bsnm{Chen}, \binits{G.}},
\bauthor{\bsnm{Povey}, \binits{D.}},
\bauthor{\bsnm{Khudanpur}, \binits{S.}}:
\bctitle{Librispeech: an {ASR} corpus based on public domain audio books}.
In: \bbtitle{Proc. IEEE International Conference on Acoustics, Speech and Signal Processing (ICASSP-2015)},
pp. \bfpage{5206}--\blpage{5210}
(\byear{2015}).
\bcomment{IEEE}
\end{bchapter}
\endbibitem

\bibitem[\protect\citeauthoryear{Peng and Harwath}{2022}]{peng2022self}
\begin{bchapter}
\bauthor{\bsnm{Peng}, \binits{P.}},
\bauthor{\bsnm{Harwath}, \binits{D.}}:
\bctitle{Self-supervised representation learning for speech using visual grounding and masked language modeling}.
In: \bbtitle{Proc. Self-Supervised Learning for Speech and Audio Processing Workshop at AAAI-2022}
(\byear{2022})
\end{bchapter}
\endbibitem

\bibitem[\protect\citeauthoryear{Khorrami et~al.}{2023}]{khorrami2023computational}
\begin{bchapter}
\bauthor{\bsnm{Khorrami}, \binits{K.}},
\bauthor{\bsnm{Cruz~Bland{\'o}n}, \binits{M.A.}},
\bauthor{\bsnm{R{\"a}s{\"a}nen}, \binits{O.}}:
\bctitle{Computational insights to acquisition of phonemes, words, and word meanings in early language: Sequential or parallel acquisition?}
In: \bbtitle{Proc. Annual Meeting of the Cognitive Science Society (CogSci-2023)},
pp. \bfpage{389}--\blpage{396}
(\byear{2023})
\end{bchapter}
\endbibitem

\bibitem[\protect\citeauthoryear{Kurumada et~al.}{2013}]{kurumada2013zipfian}
\begin{barticle}
\bauthor{\bsnm{Kurumada}, \binits{C.}},
\bauthor{\bsnm{Meylan}, \binits{S.C.}},
\bauthor{\bsnm{Frank}, \binits{M.C.}}:
\batitle{Zipfian frequency distributions facilitate word segmentation in context}.
\bjtitle{Cognition}
\bvolume{127}(\bissue{3}),
\bfpage{439}--\blpage{453}
(\byear{2013})
\end{barticle}
\endbibitem

\bibitem[\protect\citeauthoryear{Khorrami et~al.}{2023}]{khorrami2023simultaneous}
\begin{bchapter}
\bauthor{\bsnm{Khorrami}, \binits{K.}},
\bauthor{\bsnm{Cruz~Bland{\'o}n}, \binits{M.A.}},
\bauthor{\bsnm{Virtanen}, \binits{T.}},
\bauthor{\bsnm{R{\"a}s{\"a}nen}, \binits{O.}}:
\bctitle{Simultaneous or sequential training? how speech representations cooperate in a multi-task self-supervised learning system}.
In: \bbtitle{Proceedings of the 31st European Signal Processing Conference (EUSIPCO)},
pp. \bfpage{431}--\blpage{435}
(\byear{2023})
\end{bchapter}
\endbibitem

\bibitem[\protect\citeauthoryear{Dosovitskiy et~al.}{2020}]{dosovitskiy2020image}
\begin{bchapter}
\bauthor{\bsnm{Dosovitskiy}, \binits{A.}},
\bauthor{\bsnm{Beyer}, \binits{L.}},
\bauthor{\bsnm{Kolesnikov}, \binits{A.}},
\bauthor{\bsnm{Weissenborn}, \binits{D.}},
\bauthor{\bsnm{Zhai}, \binits{X.}},
\bauthor{\bsnm{Unterthiner}, \binits{T.}},
\bauthor{\bsnm{Dehghani}, \binits{M.}},
\bauthor{\bsnm{Minderer}, \binits{M.}},
\bauthor{\bsnm{Heigold}, \binits{G.}},
\bauthor{\bsnm{Gelly}, \binits{S.}}, \betal:
\bctitle{An image is worth 16x16 words: Transformers for image recognition at scale}.
In: \bbtitle{Proc. International Conference on Learning Representations (ICLR-2021)},
\bconflocation{held as an online conference}
(\byear{2020})
\end{bchapter}
\endbibitem

\bibitem[\protect\citeauthoryear{Russakovsky et~al.}{2015}]{russakovsky2015imagenet}
\begin{barticle}
\bauthor{\bsnm{Russakovsky}, \binits{O.}},
\bauthor{\bsnm{Deng}, \binits{J.}},
\bauthor{\bsnm{Su}, \binits{H.}},
\bauthor{\bsnm{Krause}, \binits{J.}},
\bauthor{\bsnm{Satheesh}, \binits{S.}},
\bauthor{\bsnm{Ma}, \binits{S.}},
\bauthor{\bsnm{Huang}, \binits{Z.}},
\bauthor{\bsnm{Karpathy}, \binits{A.}},
\bauthor{\bsnm{Khosla}, \binits{A.}},
\bauthor{\bsnm{Bernstein}, \binits{M.}}, \betal:
\batitle{Imagenet large scale visual recognition challenge}.
\bjtitle{International Journal of Computer Vision}
\bvolume{115},
\bfpage{211}--\blpage{252}
(\byear{2015})
\end{barticle}
\endbibitem

\bibitem[\protect\citeauthoryear{Harwath et~al.}{2019}]{harwath2019learning}
\begin{botherref}
\oauthor{\bsnm{Harwath}, \binits{D.}},
\oauthor{\bsnm{Hsu}, \binits{W.-N.}},
\oauthor{\bsnm{Glass}, \binits{J.}}:
Learning hierarchical discrete linguistic units from visually-grounded speech.
arXiv preprint arXiv:1911.09602
(2019)
\end{botherref}
\endbibitem

\bibitem[\protect\citeauthoryear{Ilharco et~al.}{2019}]{ilharco2019large}
\begin{bchapter}
\bauthor{\bsnm{Ilharco}, \binits{G.}},
\bauthor{\bsnm{Zhang}, \binits{Y.}},
\bauthor{\bsnm{Baldridge}, \binits{J.}}:
\bctitle{Large-scale representation learning from visually grounded untranscribed speech}.
In: \bbtitle{Proc. 23rd Conference on Computational Natural Language Learning (CoNLL)},
pp. \bfpage{55}--\blpage{65}
(\byear{2019})
\end{bchapter}
\endbibitem

\bibitem[\protect\citeauthoryear{Kenton et~al.}{2019}]{kenton2019bert}
\begin{bchapter}
\bauthor{\bsnm{Kenton}, \binits{J.D.}},
\bauthor{\bsnm{Ming-Wei}, \binits{C.}},
\bauthor{\bsnm{Toutanova}, \binits{L.K.}}:
\bctitle{{BERT}: Pre-training of deep bidirectional transformers for language understanding}.
In: \bbtitle{Proc. Annual Conference of the North American Chapter of the Association for Computational Linguistics (NAACL-HLT)},
pp. \bfpage{4171}--\blpage{4186}
(\byear{2019})
\end{bchapter}
\endbibitem

\bibitem[\protect\citeauthoryear{Dunbar et~al.}{2022}]{dunbar2022self}
\begin{barticle}
\bauthor{\bsnm{Dunbar}, \binits{E.}},
\bauthor{\bsnm{Hamilakis}, \binits{N.}},
\bauthor{\bsnm{Dupoux}, \binits{E.}}:
\batitle{Self-supervised language learning from raw audio: Lessons from the zero resource speech challenge}.
\bjtitle{IEEE Journal of Selected Topics in Signal Processing}
\bvolume{16}(\bissue{6}),
\bfpage{1211}--\blpage{1226}
(\byear{2022})
\end{barticle}
\endbibitem

\bibitem[\protect\citeauthoryear{Schatz et~al.}{2013}]{schatz2013evaluating}
\begin{bchapter}
\bauthor{\bsnm{Schatz}, \binits{T.}},
\bauthor{\bsnm{Peddinti}, \binits{V.}},
\bauthor{\bsnm{Bach}, \binits{F.}},
\bauthor{\bsnm{Jansen}, \binits{A.}},
\bauthor{\bsnm{Hermansky}, \binits{H.}},
\bauthor{\bsnm{Dupoux}, \binits{E.}}:
\bctitle{Evaluating speech features with the minimal-pair abx task: Analysis of the classical mfc/plp pipeline}.
In: \bbtitle{Proc. Annual Conference of the International Speech Communication Association (Interspeech-2013)},
pp. \bfpage{1}--\blpage{5}
(\byear{2013})
\end{bchapter}
\endbibitem

\bibitem[\protect\citeauthoryear{Schatz et~al.}{2021}]{schatz2021early}
\begin{barticle}
\bauthor{\bsnm{Schatz}, \binits{T.}},
\bauthor{\bsnm{Feldman}, \binits{N.H.}},
\bauthor{\bsnm{Goldwater}, \binits{S.}},
\bauthor{\bsnm{Cao}, \binits{X.-N.}},
\bauthor{\bsnm{Dupoux}, \binits{E.}}:
\batitle{Early phonetic learning without phonetic categories: Insights from large-scale simulations on realistic input}.
\bjtitle{Proceedings of the National Academy of Sciences}
\bvolume{118}(\bissue{7}),
\bfpage{2001844118}
(\byear{2021})
\end{barticle}
\endbibitem

\bibitem[\protect\citeauthoryear{Fenson et~al.}{2007}]{CDI}
\begin{botherref}
\oauthor{\bsnm{Fenson}, \binits{L.}},
\oauthor{\bsnm{Marchman}, \binits{V.A.}},
\oauthor{\bsnm{Thal}, \binits{D.J.}},
\oauthor{\bsnm{Dale}, \binits{P.S.}},
\oauthor{\bsnm{Reznick}, \binits{J.S.}},
\oauthor{\bsnm{Bates}, \binits{E.}}:
{MacArthur-Bates} communicative development inventories: User's guide and technical manual, 2nd edition.
Baltimore, MD: Brookes Publishing Company
(2007)
\end{botherref}
\endbibitem

\bibitem[\protect\citeauthoryear{Lavechin et~al.}{2023}]{lavechin23_interspeech}
\begin{bchapter}
\bauthor{\bsnm{Lavechin}, \binits{M.}},
\bauthor{\bsnm{Sy}, \binits{Y.}},
\bauthor{\bsnm{Titeux}, \binits{H.}},
\bauthor{\bsnm{Blandón}, \binits{M.A.C.}},
\bauthor{\bsnm{Räsänen}, \binits{O.}},
\bauthor{\bsnm{Bredin}, \binits{H.}},
\bauthor{\bsnm{Dupoux}, \binits{E.}},
\bauthor{\bsnm{Cristia}, \binits{A.}}:
\bctitle{{BabySLM: language-acquisition-friendly benchmark of self-supervised spoken language models}}.
In: \bbtitle{Proc. Annual Conference of the International Speech Communication Association (Interspeech-2023)},
pp. \bfpage{4588}--\blpage{4592}
(\byear{2023})
\end{bchapter}
\endbibitem

\bibitem[\protect\citeauthoryear{Peng et~al.}{2023}]{peng2023syllable}
\begin{bchapter}
\bauthor{\bsnm{Peng}, \binits{P.}},
\bauthor{\bsnm{Li}, \binits{S.-W.}},
\bauthor{\bsnm{Räsänen}, \binits{O.}},
\bauthor{\bsnm{Mohamed}, \binits{A.}},
\bauthor{\bsnm{Harwath}, \binits{D.}}:
\bctitle{Syllable discovery and cross-lingual generalization in a visually grounded, self-supervised speech model}.
In: \bbtitle{Proc. Annual Conference of the International Speech Communication Association (Interspeech-2023)},
pp. \bfpage{391}--\blpage{395}
(\byear{2023})
\end{bchapter}
\endbibitem

\bibitem[\protect\citeauthoryear{Tomasello}{2009}]{tomasello2009}
\begin{botherref}
\oauthor{\bsnm{Tomasello}, \binits{M.}}:
The usage-based theory of language acquisition,
pp. 69--87.
Oxford University Press
(2009)
\end{botherref}
\endbibitem

\bibitem[\protect\citeauthoryear{Lavechin et~al.}{2023}]{lavechin_struggle}
\begin{botherref}
\oauthor{\bsnm{Lavechin}, \binits{M.}},
\oauthor{\bsnm{Seyssel}, \binits{M.}},
\oauthor{\bsnm{Métais}, \binits{M.}},
\oauthor{\bsnm{Metze}, \binits{F.}},
\oauthor{\bsnm{Mohamed}, \binits{A.}},
\oauthor{\bsnm{Bredin}, \binits{H.}},
\oauthor{\bsnm{Dupoux}, \binits{E.}},
\oauthor{\bsnm{Cristia}, \binits{A.}}:
Statistical learning models of early phonetic acquisition struggle with child-centered audio data.
PsyArxiV pre-print
(2023)
\end{botherref}
\endbibitem

\bibitem[\protect\citeauthoryear{Roy et~al.}{2015}]{roy2015}
\begin{barticle}
\bauthor{\bsnm{Roy}, \binits{B.}},
\bauthor{\bsnm{Frank}, \binits{M.C.}},
\bauthor{\bsnm{deCamp}, \binits{P.}},
\bauthor{\bsnm{Roy}, \binits{D.}}:
\batitle{Predicting the birth of a spoken word}.
\bjtitle{Proceedings of the National Academy of Sciences}
\bvolume{112},
\bfpage{12663}--\blpage{12668}
(\byear{2015})
\end{barticle}
\endbibitem

\bibitem[\protect\citeauthoryear{Piaget}{1954}]{piaget2013construction}
\begin{bbook}
\bauthor{\bsnm{Piaget}, \binits{J.}}:
\bbtitle{The Construction of Reality in the Child}.
\bpublisher{New York: Basic Books}, \blocation{???}
(\byear{1954})
\end{bbook}
\endbibitem

\bibitem[\protect\citeauthoryear{Hodosh et~al.}{2013}]{hodosh2013framing}
\begin{barticle}
\bauthor{\bsnm{Hodosh}, \binits{M.}},
\bauthor{\bsnm{Young}, \binits{P.}},
\bauthor{\bsnm{Hockenmaier}, \binits{J.}}:
\batitle{Framing image description as a ranking task: Data, models and evaluation metrics}.
\bjtitle{Journal of Artificial Intelligence Research}
\bvolume{47},
\bfpage{853}--\blpage{899}
(\byear{2013})
\end{barticle}
\endbibitem

\end{thebibliography}

\clearpage
\begin{appendices}

\section{Validating model training}\label{secA}

\begin{figure}[tp]
\centering
\includegraphics[width=0.5\textwidth]{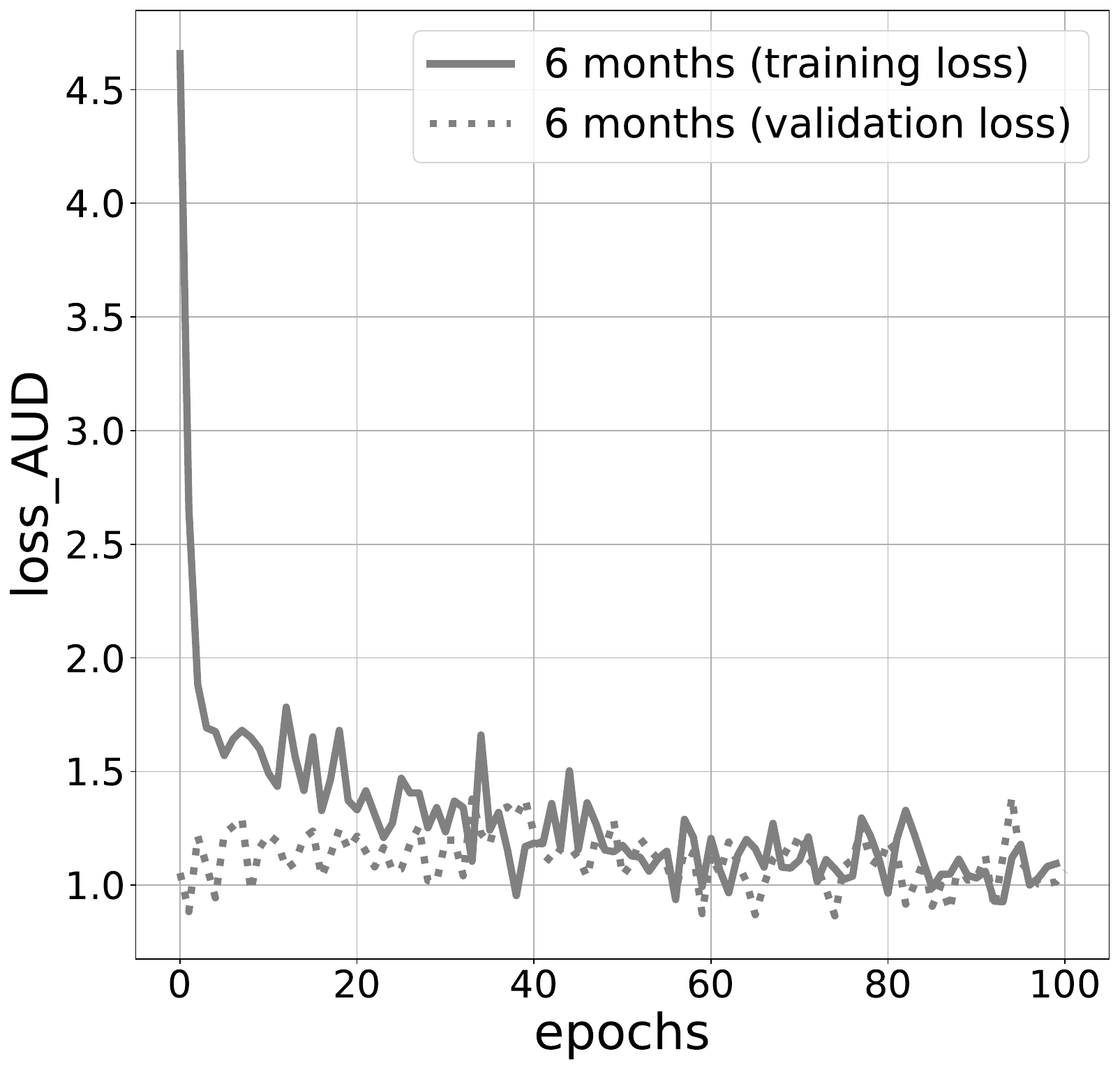}
\caption{The training and validation loss curves of the wav2vec 2.0 speech processing model ($loss_{\text{AUD}}$) over 100 epochs for the 6-month age bin. }
\label{wav2vec_fig}
\end{figure}

\begin{figure}[tp]
\centering
\includegraphics[width=0.8\textwidth]{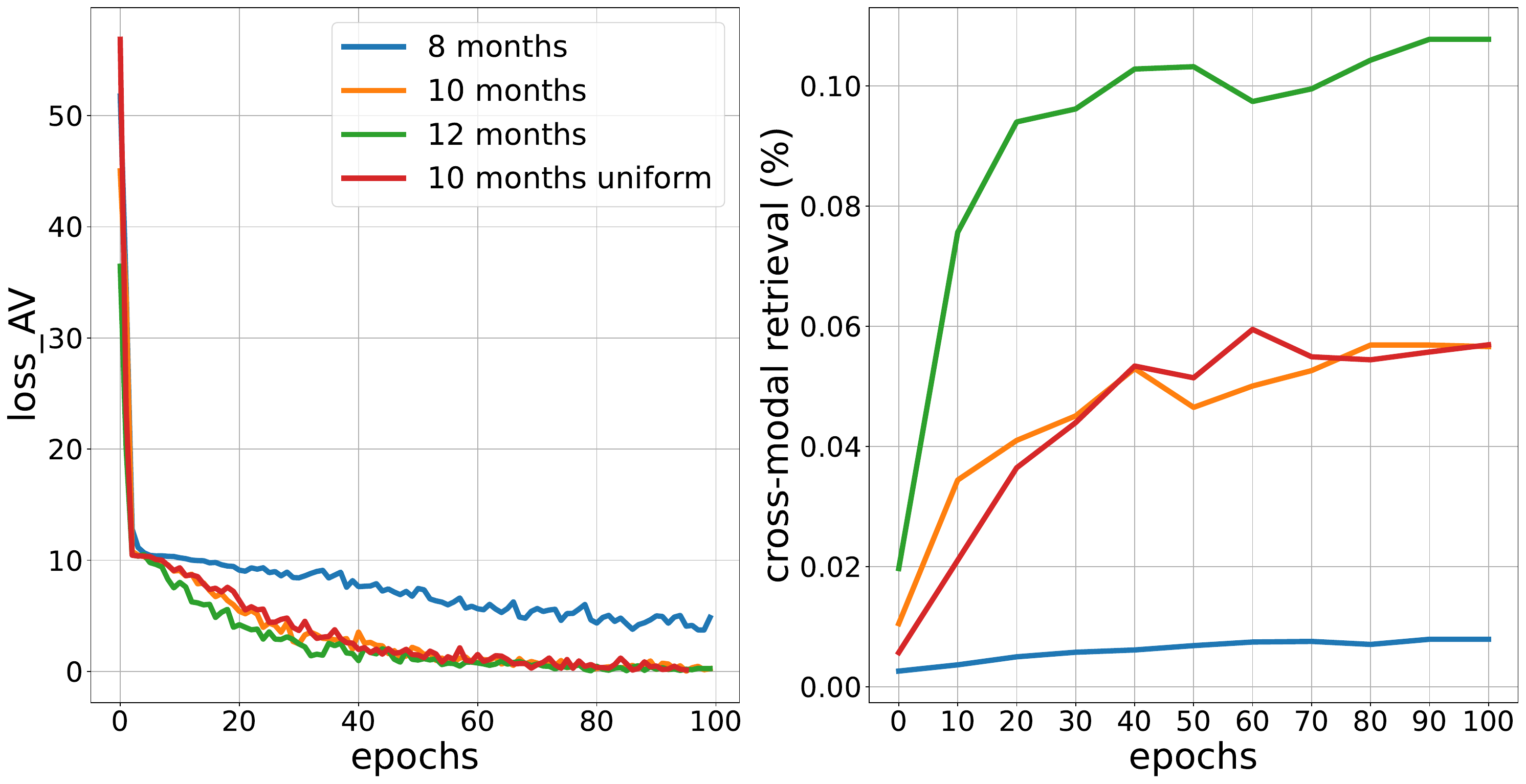}
\caption{Training statistics of the audiovisual model during the 100 epochs of training. Left panel: the mean of speech-to-image and image-to-speech recall@10 retrieval scores measured every 10 epochs. Right: The audiovisual training loss ($loss_{\text{AV}}$) of the model.}
\label{av_fig}
\end{figure}

To ensure from a machine learning perspective that the model training was successful in both the self-supervised auditory learning and during the audiovisual associative learning, we monitored the model loss functions on a separate (held-out) validation data. Fig. \ref{wav2vec_fig} shows the training and validation losses for the auditory learning using the wav2vec2.0 training procedure as a function of training epoch. Left panel of Fig. \ref{av_fig} shows the training loss for the audiovisual training in case of the four subsets of training data (2, 4, and 6 months of audiovisual learning in addition to the 4-month uniform frequency condition, all following the initial 6-months of auditory learning). Additionally, right panel in Fig. \ref{av_fig} shows the performance of the audiovisual model on speech-to-image and image-to-speech retrieval tasks by computing the recall@10 score on a set of held-out spoken utterances and image pairs. 

In the speech-to-image retrieval task, a query input consisting of a spoken caption is processed by the auditory encoder, and the resulting utterance embedding vector from the output of the auditory branch is extracted. This embedding is then compared against all possible image embedding vectors of the test dataset using a dot product, and where the image embeddings are obtained from the output of the image branch. The $k$ nearest image embeddings are then chosen. If the correct image embedding is within the set, the retrieval is considered as successful (remember that each utterance of SpokenCOCO is paired with a specific image). The recall@k metric (0--1) determines how often, on average, the correct image pair is among the nearest $k$ options.  The image-to-speech retrieval task operates similarly, assessing how often the correct speech pair of a given query image is found within the first k retrieved spoken captions. The metric, is widely employed in cross-modal semantic mapping systems \cite{hodosh2013framing} (see also \cite{chrupala2022visually}), and has been previously utilized for validating the success of audiovisual semantic learning in computational simulations of infant language acquisition \citep{khorrami2023computational, merkx2023modelling}. Overall, the plots confirm that the learning improves systematically with training in the auditory case, in the case of all audiovisual datasets. 

\clearpage

\clearpage
\section{Category-specific scores and statistics}\label{secB}

Table \ref{tab1} shows the visual category-by-category naming frequencies, average relative object size in images (object size in pixels compared to total image pixels) of the training data, and word meaning scores at 8-, 10-, and 12-months and with 10-months of training with uniform frequency distribution of naming events (for which the naming frequency per day was always 0.31).

\begin{table}[tp]
\centering
\resizebox{0.62\textwidth}{!}
{
\begin{tabular}{llrrrrrr}

\toprule
{} &         category &  freq. (per day) &   image area (\%) &   8 mo &  10 mo &  12 mo &  10 mo(u) \\
\midrule
1  &        person &         1.482 &  13.78 &  0.633 &  0.763 &  0.675 &     0.426 \\
2  &         dining table &         1.173 &  25.68 &  0.424 &  0.483 &  0.555 &     0.547 \\
3  &           dog &         1.145 &  14.68 &  0.625 &  0.644 &  0.739 &     0.759 \\
4  &         train &         1.089 &  19.59 &  0.646 &  0.741 &  0.842 &     0.595 \\
5  &        tennis racket &         1.005 &   2.08 &  0.573 &  0.857 &  0.820 &     0.788 \\
6  &           cat &         0.865 &  18.23 &  0.599 &  0.782 &  0.817 &     0.864 \\
7  &      baseball bat &         0.725 &   1.82 &  0.582 &  0.796 &  0.834 &     0.812 \\
8  &           bus &         0.725 &  24.34 &  0.799 &  0.822 &  0.856 &     0.608 \\
9  &         pizza &         0.725 &  28.13 &  0.485 &  0.960 &  0.878 &     0.839 \\
10 &    skateboard &         0.697 &   1.75 &  0.619 &  0.695 &  0.764 &     0.588 \\
11 &         horse &         0.697 &   8.20 &  0.546 &  0.554 &  0.753 &     0.520 \\
12 &           bed &         0.669 &  33.57 &  0.490 &  0.529 &  0.615 &     0.645 \\
13 &       giraffe &         0.612 &  10.46 &  0.937 &  0.939 &  0.973 &     0.983 \\
14 &           car &         0.561 &   6.84 &  0.628 &  0.660 &  0.682 &     0.787 \\
15 &          sports ball &         0.561 &   0.41 &  0.701 &  0.755 &  0.754 &     0.752 \\
16 &         clock &         0.505 &   4.48 &  0.643 &  0.731 &  0.818 &     0.655 \\
17 &        toilet &         0.505 &  10.77 &  0.537 &  0.566 &  0.901 &     0.574 \\
18 &       parking meter &         0.505 &  18.13 &  0.605 &  0.623 &  0.614 &     0.645 \\
19 &     cellphone &         0.505 &   5.60 &  0.467 &  0.473 &  0.428 &     0.430 \\
20 &    motorcycle &         0.477 &  16.12 &  0.824 &  0.888 &  0.898 &     0.906 \\
21 &          bear &         0.477 &  16.19 &  0.320 &  0.611 &  0.367 &     0.708 \\
22 &         zebra &         0.449 &  13.79 &  0.774 &  0.970 &  0.958 &     0.968 \\
23 &         bench &         0.393 &  12.09 &  0.570 &  0.721 &  0.705 &     0.724 \\
24 &      elephant &         0.393 &  18.90 &  0.686 &  0.813 &  0.934 &     0.800 \\
25 &          sink &         0.337 &   6.18 &  0.811 &  0.699 &  0.817 &     0.815 \\
26 &         truck &         0.309 &  26.46 &  0.661 &  0.856 &  0.755 &     0.753 \\
27 &          boat &         0.309 &   6.90 &  0.733 &  0.767 &  0.794 &     0.692 \\
28 &         chair &         0.309 &   5.88 &  0.623 &  0.751 &  0.761 &     0.647 \\
29 &          kite &         0.309 &   1.01 &  0.725 &  0.746 &  0.790 &     0.703 \\
30 &          cake &         0.252 &  15.48 &  0.416 &  0.305 &  0.546 &     0.485 \\
31 &          bird &         0.252 &   5.38 &  0.596 &  0.754 &  0.749 &     0.693 \\
32 &       frisbee &         0.252 &   2.75 &  0.561 &  0.768 &  0.790 &     0.675 \\
33 &          skis &         0.252 &   0.70 &  0.639 &  0.820 &  0.865 &     0.851 \\
34 &     surfboard &         0.224 &   5.26 &  0.642 &  0.683 &  0.735 &     0.610 \\
35 &           cow &         0.224 &   9.96 &  0.562 &  0.739 &  0.799 &     0.790 \\
36 &      umbrella &         0.224 &   6.68 &  0.481 &  0.512 &  0.713 &     0.596 \\
37 &        banana &         0.196 &  10.93 &  0.308 &  0.661 &  0.585 &     0.398 \\
38 &          bowl &         0.196 &  20.64 &  0.494 &  0.419 &  0.404 &     0.596 \\
39 &        laptop &         0.196 &  15.41 &  0.492 &  0.714 &  0.854 &     0.852 \\
40 &      airplane &         0.196 &  10.70 &  0.813 &  0.837 &  0.852 &     0.743 \\
41 &          vase &         0.168 &  11.53 &  0.549 &  0.573 &  0.706 &     0.623 \\
42 &     wine glass &         0.168 &   9.88 &  0.504 &  0.542 &  0.654 &     0.621 \\
43 &       traffic light &         0.168 &   2.00 &  0.590 &  0.588 &  0.837 &     0.731 \\
44 &      sandwich &         0.140 &  23.49 &  0.539 &  0.818 &  0.816 &     0.792 \\
45 &       fire hydrant &         0.140 &   9.47 &  0.521 &  0.581 &  0.707 &     0.481 \\
46 &       bicycle &         0.140 &   9.38 &  0.672 &  0.838 &  0.781 &     0.723 \\
47 &        orange &         0.140 &   6.15 &  0.258 &  0.374 &  0.544 &     0.644 \\
48 &     snowboard &         0.140 &   3.01 &  0.671 &  0.838 &  0.830 &     0.792 \\
49 &           tie &         0.112 &   4.56 &  0.543 &  0.581 &  0.632 &     0.488 \\
50 &         couch &         0.112 &  22.23 &  0.552 &  0.650 &  0.730 &     0.698 \\
51 &  refrigerator &         0.112 &  23.65 &  0.314 &  0.872 &  0.850 &     0.795 \\
52 &          book &         0.112 &   8.94 &  0.554 &  0.698 &  0.713 &     0.684 \\
53 &          stop sign &         0.084 &  11.06 &  0.491 &  0.552 &  0.620 &     0.483 \\
54 &         sheep &         0.084 &   5.64 &  0.611 &  0.826 &  0.780 &     0.571 \\
55 &           cup &         0.084 &   7.93 &  0.536 &  0.490 &  0.633 &     0.723 \\
56 &         donut &         0.084 &   8.75 &  0.657 &  0.598 &  0.726 &     0.618 \\
57 &      broccoli &         0.084 &  10.40 &  0.637 &  0.946 &  0.896 &     0.966 \\
58 &      keyboard &         0.084 &  18.79 &  0.555 &  0.671 &  0.709 &     0.779 \\
59 &            tv &         0.084 &  16.78 &  0.526 &  0.795 &  0.821 &     0.901 \\
60 &        bottle &         0.084 &   7.80 &  0.681 &  0.784 &  0.796 &     0.776 \\
61 &      suitcase &         0.056 &  17.11 &  0.480 &  0.386 &  0.644 &     0.589 \\
62 &         apple &         0.056 &   6.85 &  0.341 &  0.541 &  0.748 &     0.675 \\
63 &        carrot &         0.056 &   5.96 &  0.714 &  0.688 &  0.616 &     0.752 \\
64 &         potted plant &         0.056 &   9.71 &  0.630 &  0.824 &  0.651 &     0.801 \\
65 &      scissors &         0.056 &   5.96 &  0.550 &  0.619 &  0.667 &     0.718 \\
66 &         knife &         0.056 &   3.92 &  0.657 &  0.667 &  0.736 &     0.795 \\
67 &          oven &         0.056 &  14.84 &  0.547 &  0.587 &  0.579 &     0.646 \\
68 &         mouse &         0.056 &   2.50 &  0.484 &  0.569 &  0.642 &     0.696 \\
69 &          fork &         0.056 &   1.49 &  0.638 &  0.796 &  0.842 &     0.805 \\
70 &     microwave &         0.056 &  10.36 &  0.401 &  0.725 &  0.732 &     0.814 \\
71 &        remote &         0.056 &   9.12 &  0.501 &  0.434 &  0.370 &     0.680 \\
72 &    toothbrush &         0.028 &   3.97 &  0.529 &  0.565 &  0.643 &     0.589 \\
73 &        hot dog &         0.028 &  21.55 &  0.483 &  0.232 &  0.276 &     0.873 \\
74 &         spoon &         0.028 &   1.20 &  0.590 &  0.752 &  0.754 &     0.779 \\
75 &      backpack &         0.028 &  16.37 &  0.629 &  0.647 &  0.829 &     0.703 \\
76 &         baseball glove &         0.028 &   7.55 &  0.681 &  0.765 &  0.744 &     0.759 \\
77 &     teddy bear &         0.028 &  17.56 &  0.486 &  0.398 &  0.743 &     0.500 \\
78 &       handbag &         0.028 &  19.81 &  0.543 &  0.745 &  0.803 &     0.726 \\
79 &       toaster &         0.028 &  12.60 &  0.543 &  0.686 &  0.796 &     0.814 \\
80 &         hair drier &         0.028 &   6.62 &  0.554 &  0.611 &  0.656 &     0.654 \\
\bottomrule
\end{tabular}

}

\caption{Statistics of the 80 audiovisual categories used in the study. The 'category' column indicates the visual object category in the MSCOCO dataset. Object area is expressed as the percentage of the object area relative to the entire image, calculated for each object by averaging across all training samples. The last four columns show category-based word meaning scores for the different training age bins.}
\label{tab1}
\end{table}

\clearpage
\section{Object-word correspondences}\label{secC}

Table \ref{tab2} shows the lists of words that were considered as valid naming events of the corresponding visual object.

\begin{table}[tp]
\centering
\footnotesize
\resizebox{0.55\textwidth}{!}
{
\begin{tabular}{lll}
\toprule
{} &         category &                                              words \\
\midrule

1  &        person &                                                man \\
2  &        dining table &                                      table, tables \\
3  &           dog &                                          dog, dogs \\
4  &         train &                                      train, trains \\
5  &        tennis racket &                                             tennis \\
6  &           cat &                                          cat, cats \\
7  &      baseball bat &                                           baseball \\
8  &           bus &                                 bus, buses, busses \\
9  &         pizza &                                      pizza, pizzas \\
10 &    skateboard &  skateboard, skateboarder, skateboards, skatebo... \\
11 &         horse &                                      horse, horses \\
12 &           bed &                                          bed, beds \\
13 &       giraffe &                                  giraffe, giraffes \\
14 &           car &                                          car, cars \\
15 &          sports ball &                                        ball, balls \\
16 &         clock &                                      clock, clocks \\
17 &        toilet &                                    toilet, toilets \\
18 &       parking meter &                                      park, parking \\
19 &     cellphone &               phone, cellphone, phones, cellphones \\
20 &    motorcycle &  motorcycle, motorcycles, motorcyclist, motorcy... \\
21 &          bear &                                        bear, bears \\
22 &         zebra &                                      zebra, zebras \\
23 &         bench &                                              bench \\
24 &      elephant &                                elephant, elephants \\
25 &          sink &                                        sink, sinks \\
26 &         truck &                                      truck, trucks \\
27 &          boat &                                        boat, boats \\
28 &         chair &                                      chair, chairs \\
29 &          kite &                                        kite, kites \\
30 &          cake &                                        cake, cakes \\
31 &          bird &                                        bird, birds \\
32 &       frisbee &                                  frisbee, frisbees \\
33 &          skis &                   skis, skiers, ski, skiing, skiis \\
34 &     surfboard &                              surfboard, surfboards \\
35 &           cow &                                          cows, cow \\
36 &      umbrella &                                           umbrella \\
37 &        banana &                                    bananas, banana \\
38 &          bowl &                                        bowl, bowls \\
39 &        laptop &                                    laptop, laptops \\
40 &      airplane &                                airplane, airplanes \\
41 &          vase &                                        vase, vases \\
42 &     wine glass &                                              glass \\
43 &       traffic light &                                            traffic \\
44 &      sandwich &                               sandwich, sandwiches \\
45 &       fire hydrant &                                            hydrant \\
46 &       bicycle &                                        bike, bikes \\
47 &        orange &                                             orange \\
48 &     snowboard &   snowboard, snowboarder, snowboards, snowboarders \\
49 &           tie &                                                tie \\
50 &         couch &                                     couch, couches \\
51 &  refrigerator &                        refrigerator, refrigerators \\
52 &          book &                                        book, books \\
53 &          stop &                                               stop \\
54 &         sheep &                                              sheep \\
55 &           cup &                                          cup, cups \\
56 &         donut &                                      donuts, donut \\
57 &      broccoli &                                           broccoli \\
58 &      keyboard &                                keyboard, keyboards \\
59 &            tv &                                         television \\
60 &        bottle &                                    bottle, bottles \\
61 &      suitcase &                                suitcase, suitcases \\
62 &         apple &                                      apples, apple \\
63 &        carrot &                                    carrots, carrot \\
64 &         potted plant &                                      plant, plants \\
65 &      scissors &                                           scissors \\
66 &         knife &                                      knife, knives \\
67 &          oven &                                               oven \\
68 &         mouse &                                              mouse \\
69 &          fork &                                        fork, forks \\
70 &     microwave &                                          microwave \\
71 &        remote &                                             remote \\
72 &    toothbrush &                           toothbrush, toothbrushes \\
73 &        hot dog &                                    hotdog, hotdogs \\
74 &         spoon &                                      spoon, spoons \\
75 &      backpack &                                backpack, backpacks \\
76 &         baseball glove &                                      glove, gloves \\
77 &     teddy bear &                                              teddy \\
78 &       handbag &                                              purse \\
79 &       toaster &                                            toaster \\
80 &         hair drier &                                              dryer \\
\bottomrule
\end{tabular}
}
\caption{The object-word pairing list. Column "category" indicates the visual object category in the MSCOCO dataset. The word list comprises the set of words that were considered as valid naming events of the corresponding visual category, as they appeared in the utterances of the audiovisual training sets.}
\label{tab2}
\end{table}

\end{appendices}

\end{document}